\documentclass{article}
\usepackage{amsmath}
\usepackage[dvips]{graphicx}
\providecommand{\href}[2]{#2}   

\title{A Model for Topological Fermions}

\author{Manfried Faber\\
        Institut f\"ur Kernphysik\\
        Technische Universit\"at Wien, A--1040 Wien, Austria\\
        faber@kph.tuwien.ac.at}

\date{}      

\begin{document}

\maketitle                   

\begin{abstract}
We introduce a model designed to describe charged particles as stable topological solitons of a field with values on the internal space $S^3$. These solitons behave like particles with relativistic properties like Lorentz contraction and velocity dependence of mass. This mass is defined by the energy of the soliton. In this sense this model is a generalisation of the sine--Gordon model\footnote{We do not chase the aim to give a four--dimensional generalisation of Coleman's isomorphism between the Sine--Gordon model and the Thirring model which was shown in 2--dimensional space-time.} from 1+1 dimensions to 3+1 dimensions, from $S^1$ to $S^3$. For large distances from the center of solitons this model tends to a dual $U(1)$-theory with freely propagating electromagnetic waves. Already at the classical level it describes important effects, which usually have to be explained by quantum field theory, like particle-antiparticle annihilation and the running of the coupling.
\end{abstract}

\section{Introduction}

The standard model of particle physics describes electromagnetic, weak and strong interactions and includes three generations of fermions. Up to now all possible predictions of the standard model are in agreement with experiment. Nevertheless, in connection with the standard model there are still many interesting questions which demand for further explanations. The deepest of them may be the origin of mass, the experimental evidence for three generations of fermions and of three types of interactions, strong, electro-weak and gravitational forces. Further there are a lot of interesting properties which by the standard model are accounted for in a very abstract way. A prominent problem is the running of the couplings. The naivly formulated action turns out to predict only useless infinite numbers. A redefinition of the theory is necessary. In the renormalisation procedure it turns out that the theory can only be formulated in a consistent way if the coupling ``constant'' varies with the transfered momentum, a property which is reproduced very well in high energy experiments. In the standard model it is necessary to treat fermions with special anticommuting numbers, with Grassman variables, in order to account for the Pauli principle. The vacuum of quantum field theory turns out to be a most complicated state. The QCD vacuum, the vacuum of Quantum Chromo Dynamics, seems to account for the confinement of colour charges only, if it has the properties of a magnetic medium. In this sense quantum field theory has reintroduced the concept of ether in a form which is compatible with relativity.

There are further many well established and accepted effects which could ask for some deeper understanding. To this group belong the well known relativistic properties of particles, like the velocity dependence of mass, Lorentz contraction and time dilatation which are usually derived from the basic Lorentz transformation properties and in this way not accessible to an intuitive understanding.

There are questions which occupy the mind of physicists since decades like the Einstein-Rosen-Podolsky paradox and the absence of magnetic monopoles.

From an extreme point of view one could compare the high level of abstaction of theoretical physics with the epicycles in the Ptolemaeic world system and hope for a more intuitive explanation of physical phenomena.

There are some hints for the origin of the above mentioned problems. Gauge theories, the basic theories for the standard model, treat fundamental fermions as point-like particles. The fermionic determinant in the path integral formulation of quantum field theory turns out to be a summation over all possible pathes of these point-like fermions. The fermions interact with the gauge field only via the registration of a generalised phase of the gauge fields along these pathes. A long standing puzzle in this connection is the so called fermion doubling problem. It was proven by Nielsen and Nyonima \cite{NN81} that it is not possible to formulate fermionic fields in a discretised world with periodic boundary conditions in a unique manner.

There is a simple 1+1--dimensional model \cite{Re94}, the sine--Gordon model which has a characteristic type of $2\pi$-kink solutions which are topological excitations. They have many simple properties, which request a rather complicated treatment within the standard model of particle physics. A chain of elastically connected pendula on a horizontal mechanical transmission line is a simple mechanical model which in the continuum limit follows the sine--Gordon equation. The properties of a moving kink give a very intuitive understanding of relativistic properties. A kink-soliton can be accelerated by an increase of torsion and therefore of energy stored in the kink. ``Lorentz'' contraction and the velocity dependence of the energy of the kink follow exactly the well-known formulae of special relativity. The velocity of the kink is limited to the speed of the propagation of small amplitude waves, i.e. the mechanical analog of the speed of light. Kinks and anti-kinks behave like positive and negative charges. Equal kinks repel and opposite kinks attract each other. Kink-antikink pairs can annihilate to two waves propagating in opposite directions. The repulsion of two kinks is increasing with their relative velocity. The increase of the torsion of kinks in a collision may appear as a simple analog of the running of the coupling with the momentum transfer. Two kinks can never be at the same place -- a behaviour requested by the Pauli exclusion principle. These analogies between sine--Gordon model and basic physical properties don't include the propagation of waves. In the sine--Gordon model there are no free waves propagating with the speed of light. The potential energy allows only for photons with a mass. Low amplitude waves follow a 1+1--dimensional Klein-Gordon equation.

The analogies of the sine--Gordon model inspired Skyrme in his hope for an Unified Field Theory of Mesons and Baryons \cite{MRS93}. It seems that the investigations in this field were restricted to the description of strongly interacting particles. We will present in this article an application of similiar ideas to the description of electric charges, especially electrons, and electro-magnetic fields. The 3+1--dimensional model presented in this article tries to generalize the appealing properties of the sine--Gordon model and includes further free propagating waves.

\section{Formulation of the model}\label{Formulation}

The fundamental idea of gauge theories is the existence of a local coordinate system for the internal degrees of freedom like weak-isospin or colour and the independence of all observations from the orientation of this system. One of the basic ideas of our model is that gauge theories need only be reproduced at large distances from charges. At short distances we want to generalize the sine--Gordon model where one assigns to any point in 1+1--dimensional space-time a definite field variable, a phase. According to the standard model of strong and electro-weak interactions the most obvious internal (``colour'') degrees of freedom seem to be an $SU(2)$-field for leptons and an $SU(3)$-field for hadrons. In the following we will concentrate on $SU(2)$-fields. It is essential for our model that these fields can form topological solitons with the properties of monopoles.

To emphasize the strong relation between our model and differential geometry and in order to have a sound basis for a comparison of this model with the Skyrme model and gauge theories, we introduce in subsection \ref{LocCoordSys} the concept of local coordinate systems, connection and curvature. Since we do not intend to derive a Lagrange density from another theory, in the further subsections we are going to give arguments for the form of the Lagrangian. This is chosen in such a way that solitons with a Coulomb field are topologically stable. Readers familiar with differential geometry and gauge theories are adviced to go immediately to subsection \ref{4dimForm} where the Lagrange density is introduced which is the mathematical starting point of the model and only then to have a look at the hedgehog ansatz (\ref{hedgehog}), the energy density for static soliton fields (\ref{staticE}), the non-linear differential equation (\ref{nlDE}) and the discussion of its solutions in Sect.(\ref{MonopoleSolutions}).

\subsection{Local coordinate systems and geometry}\label{LocCoordSys}
We want to generalise the 1+1--dimensional sine--Gordon model with field variables in $S^1$ to a 3+1--dimensional model with field variables in $S^3$, the three--dimensional sphere in four dimensions. We introduce in this subsection the differential geometry on $S^3$. There are various possibilities to introduce an algebra for this space. The simplest of them may be an $SU(2)$-field in the defining representation and the oldest algebra is that of Hamiltons quaternions \cite{MTW73}. Quaternions are generalisations of complex numbers which contain a real part and three ``imaginary'' parts with the corresponding units {\bf i},  {\bf j} and {\bf k}, with the non-commutative multiplication law ${\bf i}^2 = {\bf j}^2 = {\bf k}^2 = {\bf i}{\bf j}{\bf k} =  -1$. It is well known that the properties of these units are just the properties of Paulis spin matrices ${\bf i} = -i \sigma_1$, ${\bf j} = -i \sigma_2$, ${\bf k} = -i \sigma_3$.

An $SU(2)$-field assigns to every point $x=(ct,\vec{x})$ in space-time a unimodular quaternionic number $Q(x)$ by
\begin{equation}\label{q1234}
Q = q_0 + i \sigma_k q_k = 
\left(\begin{array}{cc}
q_0 + i q_3  &  q_2 + i q_1 \\
-q_2 + i q_1  &  q_0 - i q_3
\end{array}\right) \quad \text{with} \quad  |Q^2| := q_0^2 + \vec{q}^2 = 1 .
\end{equation}

The condition $q_0^2 + \vec{q}^2 = 1$ restricts the parameter-space of $SU(2)$-matrices to the unit-sphere $S^3$ of $R^4$. The  unimodularity of $SU(2)$-matrices shows up in their exponential representation as a generalised phase-factor
\begin{equation}\label{QexpRep}
Q  =  e^{i \alpha \vec{n} \vec{\sigma}} \quad \text{with} \quad q_0 = \cos \alpha, \; q_k = n_k \sin \alpha, \; \vec{n}^2 = 1.
\end{equation}
The generalised phase $\alpha \vec{n} \vec{\sigma}= \alpha n_k \sigma_k$ is a generalised ``imaginary'' part of a quaternionic number, an element of the $su(2)$-algebra (in the defining representation).\footnote{We use the summation convention that any latin index that is repeated in a product is automatically summed on from 1 to 3. The arrows on variables in the internal space indicate the set of 3 elements $\vec{q}=(q_1, q_2, q_3)$ or $\vec{\sigma}=(\sigma_1, \sigma_2, \sigma_3)$ and $\vec{q} \vec{\sigma}= q_k \sigma_k$.}

We can define the scalar product of two quaternions $P$ and $Q$ via the scalar product of the corresponding vectors $(q_0, \vec{q})$ and $(p_0, \vec{p})$ in  $R^4$
\begin{equation}\label{scalar product}
P \cdot Q \, = \, Q \cdot P \, := \, q_0 p_0 + \vec{q}\vec{p} \, = \, \frac{1}{2} \text{Tr}(Q P^{\dagger}) \; = \; \frac{1}{2} \text{Tr}(Q^{\dagger} P) .
\end{equation}

Along a path $\vec{x}(s)$ in space-time, parametrized by a parameter $s$, the field $Q(s)$ defines, as we will discuss below, a rotation of a local coordinate system and may induce a curvature which we can relate to the electric field strength. The situation is analog to that on the curved surface of earth, which has the topology of $S^2$, of the usual sphere. At the north-pole we can define a local two--dimensional coordinate system, e.g. by the $0^\circ$ and $90^\circ$ meridians and identify them with the real unit $1$ and the imaginary unit $i$. Therefore, any vector in the tangential plane of the north-pole can be represented by a complex number $w$. A parallel transport of the coordinate system from the north-pole along meridians, characterised by the azimutal angle $\varphi$, inherits the basic system to arbitrary positions $(\vartheta, \varphi)$ on earth. They are rotated by $-\varphi$ with respect to the direction of the meridians. A parallel transport of a two--dimensional tangential vector $w$ from $(\vartheta, \varphi)$ along an arbitrary big circle results in a change of its phase by some $\Delta \phi$. The vector-field $\vec{\Gamma}$ which describes the continous rotation of the local coordinate system depends on the direction $d\vec{s}$ of movement, it is the field of connection coefficients or Christoffel symbols
\begin{equation}
w \; \stackrel{d\vec{s}}{\rightarrow} \; w^\prime \; = \; (1 - i d\phi ) \; w \; = \; (1 - i \Gamma_s ds) \; w .
\end{equation}

For the above mentioned case of an $SU(2)$-field the parameter space is $S^3$. We can define three orthogonal unit vectors at the ``north-pole'' with $Q=1$, e.g.\ the three imaginary quaternionic units or $\sigma_1$, $\sigma_2$ and $\sigma_3$. The set of vectors ($su(2)$-algebra elements) $\vec{v} \vec{\sigma} = v_k \sigma_k$ builds the tangential space at the north-pole. The product $V = \vec{v} \vec{\sigma} Q$ is an element of the tangential space at Q, since its scalar product with $Q$ according to Eq.(\ref{scalar product}) vanishes
\begin{equation}\label{tangentialVectors}
V \cdot Q = \frac{1}{2} \text{Tr}(V Q^{\dagger}) = \frac{1}{2} \text{Tr}(\vec{v} \vec{\sigma} Q Q^{\dagger}) = \frac{1}{2} v_k \text{Tr}\sigma_k = 0  .
\end{equation}
Therefore, by its action on $\sigma_k$, $Q$ inherits a local coordinate system $\sigma_k Q$ to each position on $S^3$ and also to any point in coordinate-space $R^3$.

The three vectors
\begin{equation}\label{rightadjoint}
\sigma_k^Q := \sigma_k Q
\end{equation}
form an orthonormal basis of this tangential space\footnote{
The choice of the basis vectors does not influence the relations between physical objects. Using the left adjoint basis $\sigma_k^{\prime Q} := Q \sigma_k$ instead of the right adjoint basis $\sigma_k Q$ would be completely equivalent and would lead to the same coordinate independent final results. Gauge dependent quantities like the connection $\vec{\Gamma}_s$ would of course look slightly different
\begin{equation}
\vec{\Gamma}^\prime_s = - \dot{\alpha} \vec{n} - \sin \alpha \cos \alpha \; \dot{\vec{n}} - \sin^2 \alpha \; \vec{n} \times \dot{\vec{n}} ,
\end{equation}
as it is shown in Sec.~\ref{SectU1Gauge}.}

\begin{equation}\label{orthonormVectors}
\sigma_k^Q \cdot \sigma_l^Q = \frac{1}{2} \text{Tr} \left\{ \sigma_k Q Q^\dagger \sigma_l \right\} = \frac{1}{2} \text{Tr} \left\{ \sigma_k \sigma_l \right\} = \delta_{kl} . 
\end{equation}

The derivative of $Q(s)$ along some path parametrised by $s$
\begin{equation}
\partial_s Q = \lim_{\delta s \rightarrow 0} \frac{Q(s+\delta s) -  Q(s)}{\delta s}
\end{equation}
is a vector of this tangential space at $Q(s)$. Therefore it can be written as a linear combination of the basis vectors $\sigma_k^Q$
\begin{equation}\label{derivativeQ}
\partial_s Q = i \Gamma_{sk} \sigma_k Q = i \vec{\Gamma}_s \vec{\sigma}^Q ,
\end{equation}
where we again use the vector symbol for vectors in the tangential space $\sigma_k^Q$.
With $Q(s) = \cos \alpha(s) + i \vec{\sigma} \vec{n}(s) \sin \alpha(s)$ and 
\begin{equation}\label{non-trivialConnection}
\left( \partial_s Q \right) Q^\dagger = i \vec{\sigma} \left\{ \dot{\alpha} \vec{n} + \sin \alpha \cos \alpha \; \dot{\vec{n}} - \sin^2 \alpha \; \vec{n} \times \dot{\vec{n}} \right\} = i \vec{\sigma} \vec{\Gamma}_s 
\end{equation}
we get
\begin{equation}\label{gammavector}
\vec{\Gamma}_s =  \dot{\alpha} \vec{n} + \sin \alpha \cos \alpha \; \dot{\vec{n}} - \sin^2 \alpha \; \vec{n} \times \dot{\vec{n}} ,
\end{equation}
where the dot represents the derivative with respect to $s$.\footnote{
In gauge theories expression like $\left( \partial_s Q \right) Q^\dagger$ appearing in Eqs.(\ref{derivativeQ}) and (\ref{non-trivialConnection}) are usually associated with trivial connections and zero curvature. In this respect we would like to remind that the concepts of connection fields and curvature go back to differential geometry of curved spaces by Carl Friedrich Gau/{ss}. Our fields are defined on the unit sphere $S^3$, whose Gaussian curvature is obviously one. The curvature tensor in differential geometry is designed to measure areas on curved surfaces. Therefore, it seems rather natural to allow for non-vanishing curvature tensors for fields dwelling on unit spheres. In our notation the trivial connection would read $-2 \vec{\Gamma}_s$. The factor $2$ appears due to the fact that the $SU(2)$ generators in the fundamental representation are $\sigma_k /2$, whereas in Eq. (\ref{non-trivialConnection}) there appears only $\sigma_k$. It is easy to check that for the trivial connection $\vec{\Gamma}_s^\prime = -2 \vec{\Gamma}_s$ the Maurer-Cartan equation (\ref{MaurerCartan1}) looks like a vanishing curvature tensor (\ref{CurvTens}).}
The derivatives of the basis vectors $\sigma_k^Q$ at $Q$ along some path parametrised by $s$ have components in the tangential plane at $Q$, i.e. along $\sigma_k^Q$, and a component in radial direction $Q$. These components can be found multiplying Eq.~(\ref{derivativeQ}) from the left with $\sigma_k$ and using
\begin{equation}
( \vec{\sigma} \vec{a} ) \; ( \vec{\sigma} \vec{b} ) = \vec{a} \vec{b} + i \vec{\sigma} \vec{a} \times \vec{b} .
\end{equation}
We get
\begin{equation}
\partial_s \sigma_k^Q = \epsilon_{jkl} \Gamma_{sj} \sigma_l^Q + i \Gamma_{sk} Q = i \Gamma_{skl} \sigma_l^Q + i \Gamma_{sk} Q ,
\end{equation}
or in matrix notation
\begin{equation}\label{derivBasis}
\partial_s \vec{\sigma}^Q = i \Gamma_s \vec{\sigma}^Q + i \vec{\Gamma}_s Q .
\end{equation}
In differential geometry the equivalent formulae are known as the formulae of Gau/{ss} \cite{Fab83}. The connection matrix $\Gamma_s$ describes the (passive) rotation of the unit vectors of the local coordinate system in internal space along the curve parametrised by $s$
\begin{equation}\label{Gamma}
\left( \Gamma_s \right)_{kl} = \Gamma_{skl} = \left( T_j \right)_{kl} \Gamma_{sj} \qquad \text{with} \qquad \left( T_j \right)_{kl} = -i \epsilon_{jkl}.
\end{equation}
$T_j$ are the generators in the (adjoint) $T=1-$representation of $SU(2)$ with
\begin{equation}\label{TCommutator}
\left[ T_j , T_k \right] = i \epsilon_{jkl} T_l , \qquad  T_j^\dagger = T_j , \qquad  \text{Tr}\left( T_j T_k \right) = 2 \delta_{jk} .
\end{equation}
For the understanding of the matrix elements of $\Gamma_s$ in addition to Eq.~(\ref{Gamma}) we give the relations
\begin{equation}
\Gamma_{sj} = \frac{1}{2} \text{Tr} ( T_j \Gamma_s ) , \qquad \vec{\Gamma}_s = \frac{1}{2} \text{Tr} ( \vec{T} \Gamma_s )
\end{equation}
which we use throughout the article.

From the commutatitivity of derivatives we get
\begin{equation}\label{SecondDeriv}
\begin{split}
&\frac{1}{i} \partial_s \partial_t \sigma_k^Q = \frac{1}{i} \partial_t \partial_s \sigma_k^Q = \\
& = \left( \partial_t \Gamma_{skm} + i \Gamma_{skl} \Gamma_{tlm} + i \Gamma_{sk} \Gamma_{tm} \right) \sigma_m^Q + \left( \partial_t \Gamma_{sk} + i\Gamma_{skl} \Gamma_{tl} \right) Q =\\
& = \left( \partial_s \Gamma_{tkm} + i \Gamma_{tkl} \Gamma_{slm} + i \Gamma_{tk} \Gamma_{sm} \right) \sigma_m^Q + \left( \partial_s \Gamma_{tk} + i\Gamma_{tkl} \Gamma_{sl} \right) Q .
\end{split}
\end{equation}
A comparison of the coefficients in the directions $\sigma_m^Q$ reveals the geometrical interpretation of the curvature tensor
\begin{equation}\label{CurvTensComp}
\left( R_{st} \right)_{km} = R_{stkm} = - \partial_s \Gamma_{tkm} + \partial_t \Gamma_{skm} + i \Gamma_{skl} \Gamma_{tlm} - i \Gamma_{tkl} \Gamma_{slm} ,
\end{equation}
$R$ measures the area $\vec{\Gamma}_s \times \vec{\Gamma}_t$ in internal space
\begin{equation}\label{TheoremaEgregium}
R_{stkm} = - i \left( \Gamma_{sk} \Gamma_{tm} - \Gamma_{tk} \Gamma_{sm} \right) = - i \epsilon_{lkm} \epsilon_{lij} \Gamma_{si} \Gamma_{tj} = \left( T_l \right)_{km} \left( \vec{\Gamma}_s \times \vec{\Gamma}_t \right)_l .
\end{equation}
In differential geometry $R$ is called the Riemann-Christoffel curvature tensor.

In agreement with the Theorema Egregium of Carl Friedrich Gau/{ss} Eq.~(\ref{TheoremaEgregium}) relates the area $\vec{\Gamma}_s \times \vec{\Gamma}_t$ with the connection $\Gamma$ and its derivatives. The step from differential geometry to physics is done by interpreting the curvature tensor $R$ as a field in four--dimensional space-time. Two indices $(s,t)$ of $R$ are related to the coordinate space and two indices $(k,m)$ to the internal (colour) space. $R$ has these properties in common with the field strength tensor $F$ in non-abelian gauge theories. Up to a sign convention the relation of $F$ to the gauge field $A$ is identical to the relation between $R$ and $\Gamma$. It reads in matrix notation
\begin{equation}\label{CurvTens}
R_{st} = - \partial_s \Gamma_t + \partial_t \Gamma_s + i \left[ \Gamma_s, \Gamma_t \right] .
\end{equation}
Due to Eqs.~(\ref{Gamma}) and (\ref{TCommutator}) we get
\begin{equation}\label{CommVersProd}
\vec{T}  \left( \vec{\Gamma}_s \times \vec{\Gamma}_t \right) = - i \left[ \Gamma_s, \Gamma_t \right] .
\end{equation}

From Eqs.~(\ref{CurvTensComp}) and (\ref{TheoremaEgregium}) there follows the very important relation
\begin{equation}\label{MaurerCartan1}
\partial_s \Gamma_t - \partial_t \Gamma_s = 2i \left[ \Gamma_s, \Gamma_t\right] .
\end{equation}
After removal of the generators $\vec{T}$ it reads
\begin{equation}\label{MaurerCartan2}
\partial_s \vec{\Gamma}_t - \partial_t \vec{\Gamma}_s = -2 \vec{\Gamma}_s \times \vec{\Gamma}_t .
\end{equation}
This equation is well-known under the name Maurer-Cartan structural equation. It is the necessary and sufficient condition that the $SU(2)$-field $Q$ can be reconstructed from the vector field $\Gamma_s$ \cite{MRS93}. It follows also from the $Q$-components of Eq.(\ref{SecondDeriv}).

Due to Eqs.~(\ref{CurvTens})-(\ref{MaurerCartan2}) we have now several useful expressions for the area $R_{st} \, ds \, dt$ in internal space
\begin{equation}\label{SeveralExpressionsForR}
\begin{split}
R_{st} \, ds \, dt &= \vec{T}  \left( \vec{\Gamma}_s \times \vec{\Gamma}_t \right) \, ds \, dt = ( - \partial_s \Gamma_t + \partial_t \Gamma_s + i \left[ \Gamma_s, \Gamma_t \right] ) \, ds \, dt =\\
&=  - i \left[ \Gamma_s, \Gamma_t \right] \, ds \, dt = - \frac{1}{2} ( \partial_s \Gamma_t - \partial_t \Gamma_s ) \, ds \, dt .
\end{split}
\end{equation}

After deriving connection and curvature in agreement with differential geometry we are now going to relate them to vector potential and field strength and introduce step by step the various terms in the Hamiltonian, designed to describe particles with a Coulomb field as solitons stabilized by their topological properties. Since we would like to use this Hamiltonian or the corresponding Lagrangian as the starting point of our model there is no way to derive them. The only justification of these terms is the predictive power and the simplicity of the model. Readers which do not like such pedagogical introductions and prefer to start from a defining Lagrangian and to derive immediately all properties from it are adviced to switch to subsection~\ref{4dimForm} where the defining Lagrangian of the model is given. In subsection~\ref{SecElecMonop} we introduce the electric field and its energy. By adding  a potential energy we get in subsection~\ref{stabilization} the total energy of static field configurations. The variation of the profile function of hedgehog solutions leads to a non-linear differential equation for stable soliton configurations. We are able to solve this equation for a special case analytically in subsection~\ref{MonopoleSolutions}. Further solutions of the variational problem we investigate numerically. After introducing the magnetic field and solitons moving with constant velocity in subsection~\ref{MovingSolitons} we arrive at the space-time formulation in subsection~\ref{4dimForm}.

\subsection{The electric monopole}\label{SecElecMonop}

For the description of electric charges we want to take advantage of the topology of $S^3$ using the well-known hedgehog ansatz
\begin{equation}\label{hedgehog}
Q \; = \; \cos \alpha + i \;  n_k \sigma_k \sin \alpha, \quad \text{with} \quad \alpha = \alpha(r), \quad \vec{n}  = \frac{\vec{r}}{r}
\end{equation}
for the $SU(2)$-field $Q$ of Eq.(\ref{QexpRep}), which we would like to call the soliton-field. The parameter space of $SU(2)$ is $S^3$, the unit sphere in four dimensions. The basic idea is to describe charged particles by field configurations which cover only an odd number $n_w$ of hemispheres $S^3_{1/2}$ of $S^3$. In $S^3$ such field configurations have an open boundary $S^2_{int}$, a unit sphere in three dimensions, defined by $q_0=0, \; q_k=n_k$. This sphere in internal space is related to the infinite far sphere $S^2_{\infty}$ in coordinate space by $\vec{q}  = \vec{n} = \frac{\vec{r}}{r}$. By this construction we expect forces of infinite range between such particles. For an even number $n_w$ of coverings, $S^2_{\infty}$ is related to a single element in parameter space. We anticipate forces of finite range only.

We get regular hedgehog fields (\ref{hedgehog}) for
\begin{equation}\label{WindingNumber}
\alpha(0) = n_0 \pi , \qquad  \alpha(\infty) = \alpha(0) + \frac{n_w \pi}{2} ,
\end{equation}
and expect
\begin{equation}
\begin{align} \label{charged leptons} 
&n_w \; \cdots \; \text{odd} &&\text{charged field configuration} , \\
&n_w \; \cdots \; \text{even} &&\text{uncharged field configuration} .
\label{uncharged}
\end{align}
\end{equation}
Without loss of generality we can put $n_0 = 0$.

We aim now to relate the differential geometrical concepts of subsection~\ref{LocCoordSys} to physics and apply them to the hedgehog ansatz. We follow the concept of gauge theories and relate vector potential to connection and field strength to curvature.

In spherical coordinates the connections $\Gamma_r dr$, $\Gamma_\varphi d\varphi$ and $\Gamma_\vartheta d\vartheta$ correspond to translations by $dr$, $r d\vartheta$ and $r \sin \vartheta d\varphi$. Therefore, we introduce the differential line elements
\begin{equation}
l_r \, dr = dr, \quad
l_\vartheta \, d\vartheta = r \, d\vartheta, \quad
l_\varphi \, d\varphi = r \sin \vartheta \, d\varphi
\end{equation}
and identify up to a measure system dependent constant the curvature $R$ of Eq.~(\ref{SeveralExpressionsForR}) with the electric flux through the corresponding area in space. Since we would like to formulate a model for the most elementary charge, the electron, we write
\begin{equation}\label{EFieldStrength}
E_r = - \frac{e_0}{4\pi \varepsilon_0} \frac{R_{\vartheta\varphi}}{l_\vartheta l_\varphi}, \quad
E_\vartheta = - \frac{e_0}{4\pi \varepsilon_0} \frac{R_{\varphi r}}{l_\varphi l_r}, \quad
E_\varphi = - \frac{e_0}{4\pi \varepsilon_0} \frac{R_{r \vartheta}}{l_r l_\vartheta}.
\end{equation}
where the constant is adjustet to the charge $-e_0$ of the electron and the international measure system (SI). The connection $\Gamma$ we related correspondingly to the dual gauge field $C$ by line elements
\begin{equation}
C_r = - \frac{e_0}{4\pi \varepsilon_0} \frac{\Gamma_r}{l_r}, \quad
C_\vartheta = - \frac{e_0}{4\pi \varepsilon_0} \frac{\Gamma_\vartheta}{l_\vartheta}, \quad
C_\varphi = - \frac{e_0}{4\pi \varepsilon_0} \frac{\Gamma_\varphi}{l_\varphi}.
\end{equation}

According Eq.(\ref{gammavector}) using the hedgehog ansatz (\ref{hedgehog}) we get the connection
\begin{equation}\label{sphericalVecPots}
\begin{split}
\vec{\Gamma}_r &= \alpha^\prime(r) \, \vec{e}_r ,\\
\vec{\Gamma}_\vartheta &= \sin \alpha \left[ \cos \alpha \, \vec{e}_\vartheta - \sin \alpha \, \vec{e}_\varphi \right] = \sin \alpha \, \vec{e}_\xi ,\\
\vec{\Gamma}_\varphi &= \sin\vartheta \sin \alpha \left[ \sin \alpha \, \vec{e}_\vartheta + \cos \alpha \,  \vec{e}_\varphi \right] = \sin\vartheta \sin \alpha \, \vec{e}_\varphi ,
\end{split}
\end{equation}
where $\vec{e}_r, \vec{e}_\vartheta, \vec{e}_\varphi$ and $
\vec{e}_\xi = \cos \alpha \; \vec{e}_\vartheta - \sin \alpha \; \vec{e}_\varphi$, 
$\vec{e}_\eta =  \sin \alpha \; \vec{e}_\vartheta + \cos \alpha \;  \vec{e}_\varphi$ are the corresponding unit vectors and the prime indicates derivatives with respect to $r$. 

The electric field strength which is defined up to the measure dependent constant as the ratio of the area $R$ in internal space to the corresponding area in physical space reads with Eq.~(\ref{SeveralExpressionsForR})
\begin{equation}\label{EFieldStrengthHedgeHog}
\begin{split}
&E_r = - \frac{e_0}{4\pi \varepsilon_0} \frac{( \vec{\Gamma}_\vartheta \times \vec{\Gamma}_\varphi ) \vec{T}}{r^2 \sin \vartheta} = - \frac{e_0}{4\pi \varepsilon_0} \frac{\sin^2 \alpha}{r^2} \, \vec{T} \vec{e}_r := E_r^r \, T_r, \\
&E_\vartheta = - \frac{e_0}{4\pi \varepsilon_0} \frac{( \vec{\Gamma}_\varphi \times \vec{\Gamma}_r ) \vec{T}}{r \sin \vartheta} = - \frac{e_0}{4\pi \varepsilon_0} \; \frac{\alpha^\prime(r) \, \sin \alpha}{r} \, \vec{T} \vec{e}_\xi := E_\vartheta^\xi \, T_\xi,\\
&E_\varphi = - \frac{e_0}{4\pi \varepsilon_0} \frac{( \vec{\Gamma}_r \times \vec{\Gamma}_\vartheta ) \vec{T}}{r} = - \frac{e_0}{4\pi \varepsilon_0}\; \frac{\alpha^\prime(r) \, \sin \alpha}{r} \, \vec{T} \vec{e}_\eta := E_\varphi^\eta \, T_\eta  .
\end{split}
\end{equation}

These expressions correspond to the field strength tensors in dual non-abelian gauge theories. If $\alpha^\prime(r)$ approaches zero fast enough at infinity the behaviour at large distances is dominated by the curvature component $R_{\vartheta\varphi}$. In this case we get for odd winding number $n_w$ a field strength vanishing like $1/r^2$. The electric flux through the infinite far sphere  $S^2_{\infty}$ remains finite
\begin{equation}
\oint_{S^2_{\infty}} d\vec{S} \vec{E} = - \, \frac{e_0}{\varepsilon_0} \, T_r
\end{equation}
and is given by the charge of the electron in agreement with Gau/{ss}'s law.

In the Skyrme model, which in our notation treats only even winding numbers $n_w$, the field strength vanishes faster than $1/r^2$ at spatial infinity. In that model it is common to use another definition of the field strength. It reads in our notation\footnote{$F_{\mu \nu}$ is here the ordinary field strength tensor of electrodynamics in analogy to the tensor in Eq.~(\ref{dualTrafo}) }
\begin{equation}\label{SkyrmeFieldStrength}
F_{s t} \; \propto \; \partial_s \Gamma_t - \partial_t \Gamma_s - 2 i \left[ \Gamma_s, \Gamma_t\right] = 0 .
\end{equation}
This expression vanishes due the Maurer-Cartan equation (\ref{MaurerCartan1}), see e.g. Eq. (1.34) of ref. \cite{MRS93}. Since this field strength is zero, this definition would suppress the observation of the electric field produced by a soliton at large distances. In this case the soliton could not be interpreted as a physical electron whose electric field behaves like a Coulomb field. That is why in our case the field strength and the curvature related to it should be defined in a way which is asymptotically compatible with the field strength of an electromagnetic field at large distances.

In this case at $r \rightarrow \infty$ only the radial part of the electric field should survive, and the observer should feel only the Coulomb--like field of an electric charge. Therewith the observer would not be able to distinguish whether this field is produced by a point--like particle or an extended topological object. Since experimentally our topological object should be detected by virtue of its electric charge we prefer to call this topological object an electron.

The contribution of the electric field to the energy density we define according the non-abelian gauge theories by
\begin{equation}\label{electric density}
{\cal H}_e \; := \; \frac{\varepsilon_0}{4} \text{Tr} \; \vec{E}^2 .
\end{equation}
For the energy of the electric field (\ref{EFieldStrengthHedgeHog}) follows
\begin{equation}\label{electric energy}
\begin{split}
H_e \; &= \; 
\int{d^3 r} {\cal H}_e \; = \; 
\frac{\varepsilon_0}{2} \int{d^3 r} \left[ (E_r^r)^2 + (E_\vartheta^\xi)^2 + (E_\varphi^\eta)^2 \right] \; = \;\\ 
&= \;\alpha_f \hbar c \int_0^\infty{dr} \left[ \frac{\sin^4 \alpha}{2 r^2} + (\cos^\prime \alpha)^2 \right] ,
\end{split}
\end{equation}
with Sommerfeld's finestructure constant $\alpha_f = \frac{e_0^2}{4 \pi \varepsilon_0 \hbar c }$ and the metric tensor $\eta_{\mu\nu} = \text{diag}(1,-1,-1,-1)$.

In the Skyrme model, see e.g. \cite{MRS93}, the energy contribution (\ref{electric density}) is called the ``Skyrme term''. It is very well known that the Skyrme term is not able to stabilize a topological field configuration. This can be concluded from a simple scaling argument. Lets start from an arbitrary field configuration with energy $E = H_e$ and make a scale transformation by a factor $\lambda > 1$, it follows, the energy decreases by a factor $1/\lambda$ and the minimisation of the integral in (\ref{electric energy}) leads to an unbounded increase of the size of the object.

\subsection{Stabilization of the monopole solution}\label{stabilization}

In order to stabilize the solution (\ref{charged leptons}) we have to add a term which scales under transformations $\vec{r} \rightarrow \lambda \vec{r}$ with a positive power of $\lambda$. If we used a mass term for the vector fields $C$ we would arrive at the Skyrme model which has only topological stable solutions of type (\ref{uncharged}).

The ``potential'' term which we look for should vanish for $\alpha = (2n+1) \pi / 2$ and should be an even function of $q_0 = \cos \alpha$. We will discuss in the following ``potential'' energy densities of the form
\begin{equation}\label{potential}
{\cal H}_p \; = \; 
\frac{\alpha_f \hbar c}{4 \pi r_0^4} \left( \frac{\text{Tr} Q}{2} \right)^{2m} \; = \; 
\frac{\alpha_f \hbar c}{4 \pi r_0^4} \cos^{2m} \alpha , \quad m=1,2,3,\cdots ,
\end{equation}
where by dimensional arguments we had to introduce a fundamental length parameter $r_0$. Such potential terms scale with $\lambda^3$ and keep the topological soliton together like the potential energy term in our one--dimensional analogon, the sine--Gordon model.

With this assumption the total energy functional reads in scaled coordinates $\rho =\frac{r}{r_0}$
\begin{equation}\label{staticE}
H \; = \; H_e + H_p \; = \; \frac{\alpha_f \hbar c}{r_0} \int_0^\infty{d\rho} \left[ \frac{\sin^4 \alpha}{2 \rho^2} + (\partial_\rho \cos \alpha)^2  + \rho^2 \cos^{2m} \alpha \right] .
\end{equation}
which besides the power $m$ contains a single free parameter, a fundamental length scale $r_0$. Another possibility would be to use  $\cot^2 \alpha$ in the potential energy density.

All contributions to the static energy (\ref{staticE}) are positive. The energy is therefore bounded from below by the energy of trivial field configurations with $Q=$const. and $q_0 = \cos \alpha = 0$. Every topological sector, characterised by its winding number $n_w$ which was defined in Eq.~(\ref{WindingNumber}), has its own minimal configuration. There is a simple reason for the stability of non-trivial field configurations, the competition between two terms in the energy with different scale dependence. The energy density of the electric field tries to diminish the curvature of the field configuration and therefore to enlarge the region of space with $\cos \alpha \ne 0$. On the other hand the potential energy aims at compressing the field configuration.

The existence of stable field configurations follows from the solution of the variational problem. Variation with respect to $\cos \alpha(\rho)$ in Eq.~(\ref{staticE}) leads to a non-linear second order differential equation
\begin{equation}\label{nlDE}
\partial^2_\rho \cos \alpha \; + \; \frac{(1 - \cos^2 \alpha) \cos \alpha}{\rho^2} \; - \; m  \rho^2 \cos^{2m-1} \alpha \; = \; 0 
\end{equation}
with the boundary condition (\ref{WindingNumber}) for the hedgehog ansatz (\ref{hedgehog}). Some of the solutions of (\ref{nlDE}) we will discuss in the next subsection.

Stabilization is a common problem for models of topological objects. For the stabilization of the topological solution in the Georgi-Glashow model \cite{GG72}, based on a triplet of scalar Higgs fields and the $SO(3)$-gauge group, 't Hooft and Polyakov \cite{tHP74} used a Higgs-potential. Of course, due to the consideration of classical solutions of equations of motion they could have used any potential for scalar fields leading to the Higgs mechanism providing the appearence of non-zero masses of gauge fields in the spontaneously broken phase. The standard form of the Higgs-potential for the stabilization of the soliton is imposed by the requirement of renormalisability. If one is not restricted by the requirement of renormalisability, that is just the case of our approach, the stabilizing potential can be any one \cite{BM93}. Another important point distinguishing our approach and that of 't Hooft--Polyakov is in the definition of the field strength. In the 't Hooft--Polyakov approach the Higgs field is used as a lens projecting the non-abelian field strength onto the electromagnetic one. In our case such a projection runs automatically through the steep fall-off of the transversal components of the electric field of the topological electron at large distances. Concerning the renormalisability of the model preliminary results obtained for the Hamiltonian induced by quantum fluctuations around static topologigal solitons in the hedgehog ansatz (\ref{hedgehog}) have shown the appearance of only trivial divergencies which can be absorbed in the normalisation factor of the partition function \cite{FI00}.

\subsection{Discussion of monopole solutions}\label{MonopoleSolutions}

\begin{figure}
\begin{center}
\parbox{5cm}{\centerline{\scalebox{0.5}{\includegraphics{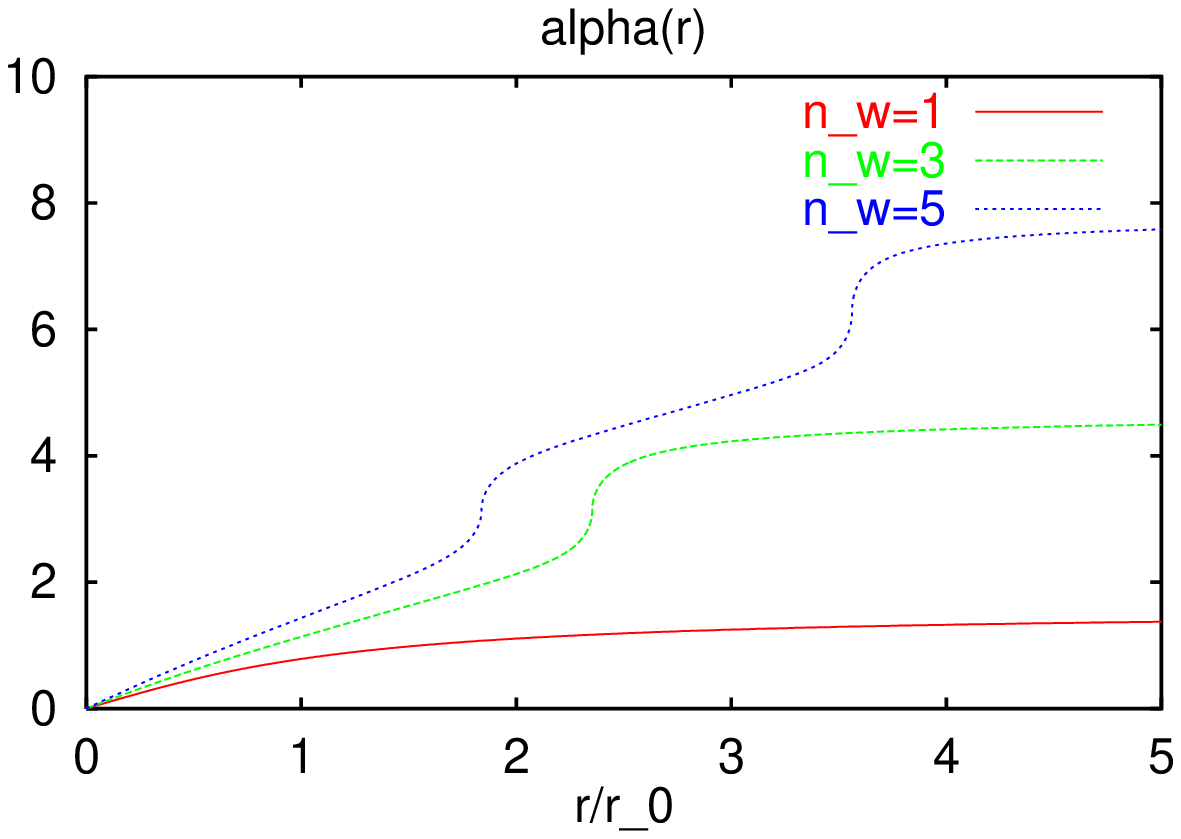}}}}
\hspace{1cm}
\parbox{5cm}{\centerline{\scalebox{0.5}{\includegraphics{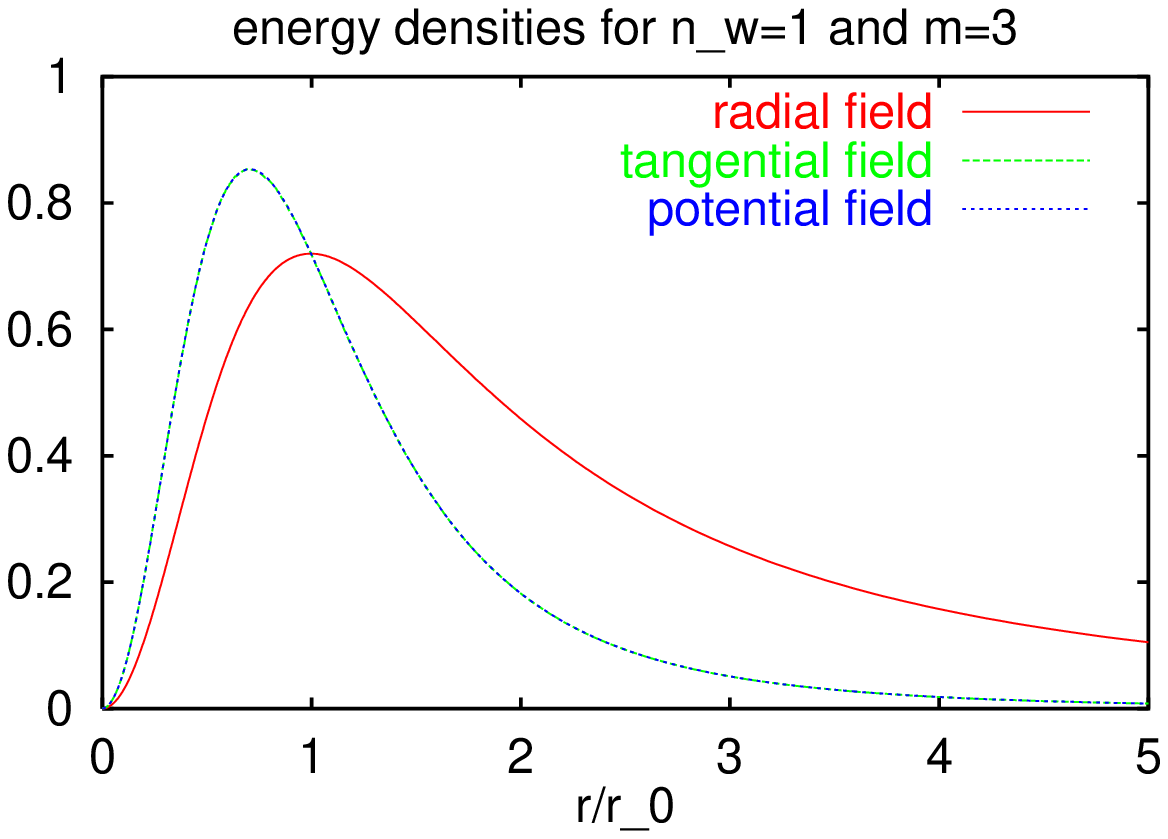}}}}\\
\parbox{5cm}{\centerline{\scalebox{0.5}{\includegraphics{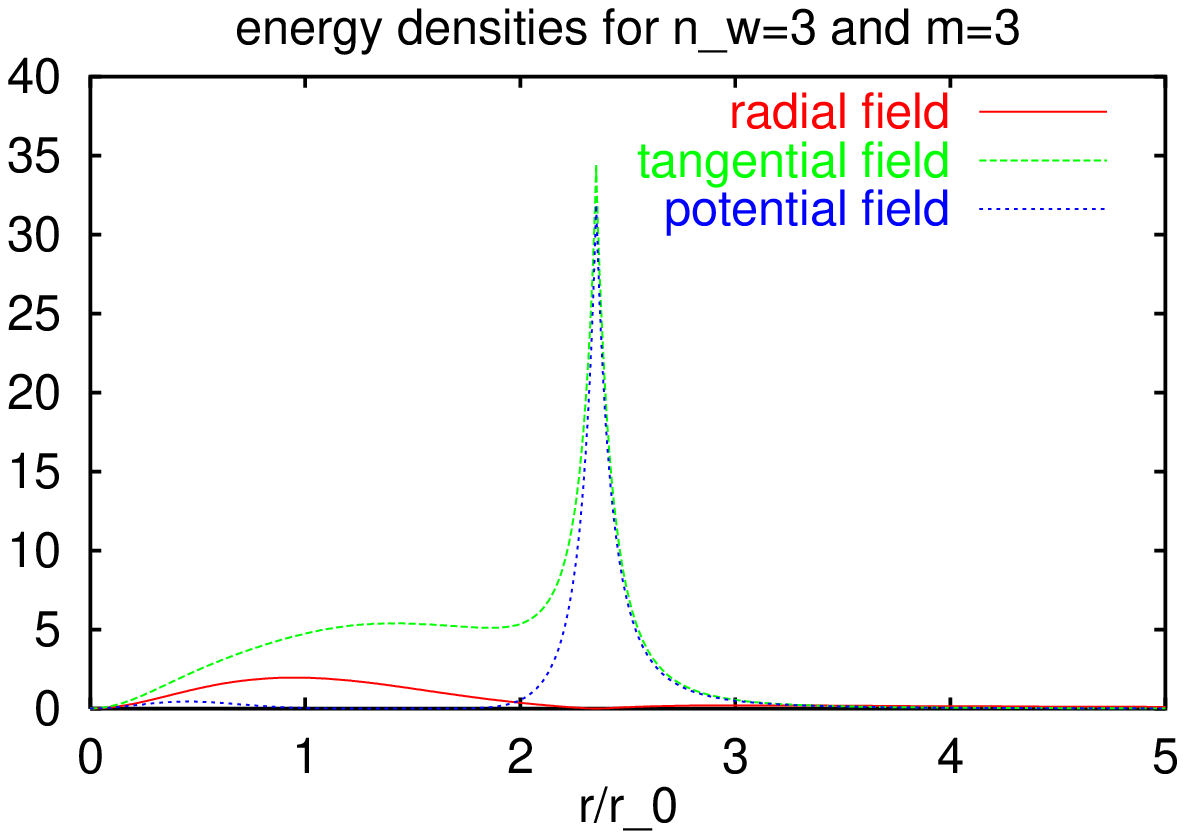}}}}
\hspace{1cm}
\parbox{5cm}{\centerline{\scalebox{0.5}{\includegraphics{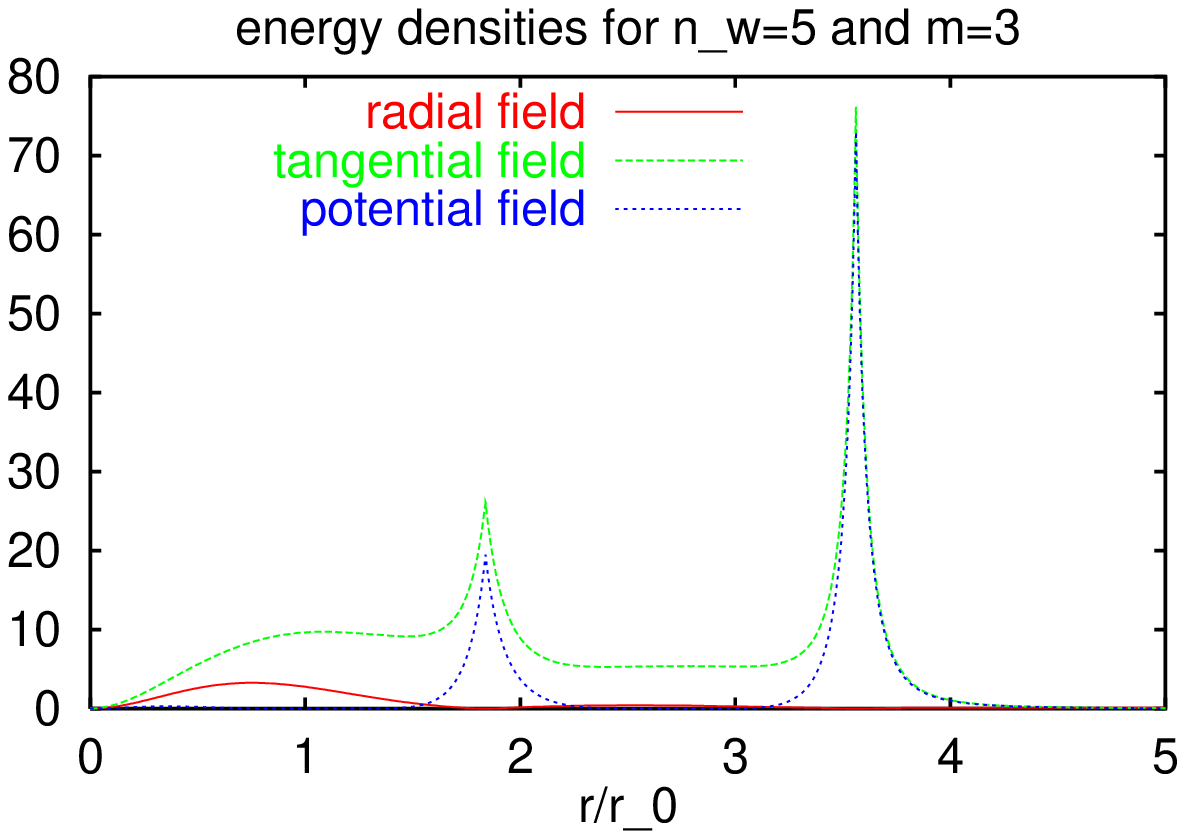}}}}
\end{center}
\caption{Profile function $\alpha(r)$ and the three contributions to the radial energy densities  $h = 4 \pi r^2 {\cal H}$ according to Eq.~(\ref{staticE}) in units of $\frac{\alpha_f \hbar c}{r_0}$ as a function of $\frac{r}{r_0}$ for $n_w = 1, \; 3$ and $5$, for a potential energy term with power $m=3$. According to Eq.~(\ref{radialdensities}) the tangential energy density and the potential energy density coincide for $n_w=1$.}
\label{enedensvonr}
\end{figure}

In Eq.(\ref{staticE}) we gave the energy of a static field configuration of hedgehog type (\ref{hedgehog}). We have shown that the minimisation of this energy leads to the non-linear differential equation (\ref{nlDE}). For $m=3$ we can give the lowest energy solution ($n_w = 1$) of this minimisation problem in analytical form
\begin{equation}
\alpha(r) \; = \; \text{atan}(\rho) \quad \text{with} \quad \rho = \frac{r}{r_0} .
\end{equation}
It follows
\begin{equation}
\sin \alpha(r) \; = \; \frac{\rho}{\sqrt{1+\rho^2}} , \quad \cos \alpha(r) \; = \; \frac{1}{\sqrt{1+\rho^2}} ,
\end{equation}
with the radial energy densities $h = 4 \pi r^2 {\cal H}$
\begin{equation}\label{radialdensities}
\begin{split}
h_{rad} \; &= \; \frac{\alpha_f \hbar c}{r_0} \frac{\rho^2}{2(1+\rho^2)^2} , \\
h_{tan} \; &= \; \frac{\alpha_f \hbar c}{r_0} \frac{\rho^2}{(1+\rho^2)^3} , \\
h_p \; &= \; \frac{\alpha_f \hbar c}{r_0} \frac{\rho^2}{(1+\rho^2)^3} ,
\end{split}
\end{equation}
for the radial field, the tangential field and the potential term, respectively. The profile function $\alpha(r)$ and the radial dependence of the radial energy densities $h$ are shown in the upper two diagrams of Fig.~\ref{enedensvonr}.

Their integrals behave like 2:1:1 and sum up to the total energy
\begin{equation}
H_1 \; = \; \frac{\alpha_f \hbar c}{r_0} \frac{\pi}{4} \quad \text{with} \quad \alpha_f \hbar c = 1.44 \text{~MeV~fm} .
\end{equation}
If we compare this expression with the energy of the electron of 0.511 MeV, we get $r_0 = 2.21$ fm. A value which seems rather large.

There are further solutions of Eq.~(\ref{nlDE}) with higher winding numbers $n_w$, these can be determined numerically by minimising the energy functional (\ref{staticE}). The profile function $\alpha(r)$ and the three contributions to the radial energy densities $h$ for the numerical solutions to the winding numbers $n_w = 3$ and $5$ for a potential energy term with power $m=3$ are also shown in Fig.~\ref{enedensvonr}.

\begin{figure}
\centerline{\scalebox{1.0}{\includegraphics{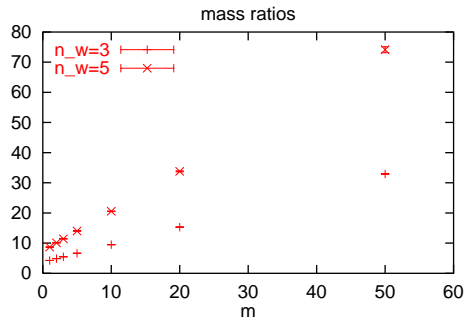}}}
\caption{Mass ratios of the mimina of the action functional (\ref{staticE}) for $n_w=3$ and $n_w=5$ compared to the solution with $n_w=1$ for some values of the power $m$ of the potential term.}
\label{massratios}
\end{figure}

For these solutions we get the following energy ratios
\begin{equation}
H_1 \; : \; H_3 \; : \; H_5 \; = \; 1 \; : \; 5.5 \; : \; 11.4 \quad \text{for} \quad m=3 .
\end{equation}
Increasing the power $m$ of the potential term (\ref{potential}) one gets higher ratios. The results for some powers are shown in Fig.~\ref{massratios}. They are again determined by numerical minimisation of the action functional (\ref{staticE}). The indicated errors are the energy differences which result from a doubling of the integration intervals. As we will discuss in more detail in Sect.~\ref{TopologicalCharge} $n_w$ is a topological quantum number which is conserved in transitions. Increasing the power $m$ the mass ratios increase with an increasing power of $n_w$ and may even reach the experimental mass ratios $1:207:3477$ of electron, muon and tauon around $m \approx 350$.
There is a certain freedom in the choice of the potential energy term (\ref{potential}). But more detailed investigations are needed in order to study all consequences of various choices of the potential.

The stable soliton solutions show an interesting relation between the contributions of the potential and the electric energy to the total energy. This relation can even be used in numerical calculations to test the accuracy of the approximation. In the next section we treat the relativistic behaviour of the solitons. It will turn out that this relation is a necessary condition for the relativistic behaviour of the solitons of our theory. One can easily show that this type of virial theorem is guaranteed by the functional form of the energy expression, it is well-known in Skyrme-models under the name of Hobart-Derrick necessity condition \cite{MRS93}. Let us follow this important argument and return to the scale transformations which we mentioned in connection with the stability of solitons. We have found that against such scale transformations $x \rightarrow \lambda x$ the electric energy behaves like $\frac{1}{\lambda}$ and the potential energy like $\lambda^3$. For a stable solution the derivative of the energy (\ref{staticE}) with respect to $\lambda$ has to vanish
\begin{equation}
\frac{d}{d\lambda} H \Bigr\rvert_{\lambda=1} \; = \; \frac{d}{d\lambda} H_e \Bigr\rvert_{\lambda=1} + \frac{d}{d\lambda} H_p \Bigr\rvert_{\lambda=1} \; = \; 3 H_p - H_e\; = \; 0 .
\end{equation}
It follows that the potential energy of a minimal configuration is one third of the static electric energy contribution. That's exactly the relation we were looking for.

\subsection{Particle properties of solitons}\label{MovingSolitons}

One of the nice features of the sine--Gordon model is the relativistic behaviour of its solitons. To gain insight into the corpuscular nature of the solitions of this model we introduce the speed of light $c$ and investigate a configuration moving with a constant velocity $\beta=v/c$ in z-direction
\begin{equation}\label{movingSoliton}
Q_v(x) = \cos \alpha(|\vec{r}-\vec{v}t|) + i \; \vec{\sigma} \frac{\vec{r}-\vec{v}t}{|\vec{r}-\vec{v}t|} \sin \alpha(|\vec{r}-\vec{v}t|) ,
\end{equation}
in four--dimensional space-time $x^\mu = (ct, \vec{x})$. For the vector-field  $\Gamma_\mu$ of connection-coefficients we define an additional time component $(\mu = 0 )$
\begin{equation}\label{fourdimVecPot}
\Gamma_\mu = \vec{T} \vec{\Gamma}_\mu = \vec{T} \frac{1}{2i} \text{Tr} \left\{ \vec{\sigma} \left( \partial_\mu Q \right) Q^\dagger \right\} .
\end{equation}
This allows to introduce a curvature in space-time directions and the corresponding physical quantity, the magnetic induction $\vec{B}$
\begin{equation}\label{MFieldStrength}
c B_k := \frac{e_0}{4\pi \varepsilon_0} \left( \partial_0 \Gamma_k - \partial_k \Gamma_0 - i \left[ \Gamma_0 , \Gamma_k \right] \right) .
\end{equation}
In addition to the static contribution to the energy density of the moving soliton we therefore expect dynamic, magnetic contributions
\begin{equation}\label{energy contributions}
\begin{split}
{\cal H} \; &:= \; {\cal T} + {\cal V} , \quad {\cal V} \; := \; {\cal H}_{e} + {\cal H}_{p} , \quad {\cal T} \; := \; {\cal H}_{m} ,\\
{\cal H}_{e} \; &:= \; \frac{\varepsilon_0}{4} \text{Tr} \vec{E}^2 , \quad {\cal H}_{m} \; := \; \frac{\varepsilon_0}{4} \text{Tr} c^2 \vec{B}^2 , \quad {\cal H}_{p} \; := \; \frac{\alpha_f \hbar c}{4 \pi r_0^4} \left( \frac{\text{Tr} Q}{2} \right)^{2m} .
\end{split}
\end{equation}

In all dynamical systems with a competition between static and dynamic degrees of freedom, allowing for an oscillatory behaviour, one has to search for solutions where variations of the static part are compensated by variations of the dynamic part and therefore the sum of both contributions is a constant of motion and the difference an extremum. This requirement is the basis for the Lagrangian formalism, where one looks for an extremum of the action
\begin{equation}\label{action}
S \; := \; \int{d^4x} {\cal L} , \quad {\cal L} \; := \; {\cal T} - {\cal V} .
\end{equation}

Due to the special $t$-dependence in Eq.~(\ref{movingSoliton}) we get
\begin{equation}
\partial_0 \; = \; - \beta \partial_z , \quad \Gamma_0 \; = \; - \beta \Gamma_z .
\end{equation}
and the relations
\begin{equation}\label{BeqBetaXE}
c B_x \; = \; - \beta E_y , \quad c B_y \; = \; \beta E_x , \quad c B_z \; = \; 0 .
\end{equation}
follow immediately. These equations are well known from special relativity for the relation between electric and magnetic fields in a reference frame moving with velocity $v$ in negative z-direction with respect to a static charge. Here, they follow from the assumption that the soliton doesn't change its shape with time.

Moving sine--Gordon solitons are compressed relative to the static configurations as requested by their relativistic behaviour. To find an analogous behaviour of our solitons we transform the coordinates $\vec{x}$ of the laboratory system to uncontracted comoving coordinates $\stackrel{\circ}{\vec{x}}$
\begin{equation}\label{Lorentz transformation}
\stackrel{\circ}{z} \; = \; \gamma (z - vt) , \qquad \stackrel{\circ}{x},\stackrel{\circ}{y} \; = \; x,y ,
\end{equation}
with a scale factor $\gamma$ which we can fix by the above mentioned variational principle. According to  (\ref{movingSoliton}) and (\ref{fourdimVecPot}) this transformation implies
\begin{equation}
\begin{split}
&E_x = \gamma \stackrel{\circ}{E}_x , \quad \quad E_y = \gamma \stackrel{\circ}{E}_y , \quad \quad E_z = \stackrel{\circ}{E}_z , \\
&c B_x = - \beta \gamma \stackrel{\circ}{E}_y , \quad c B_y = \beta \gamma \stackrel{\circ}{E}_x , \quad c B_z = 0 .
\end{split}
\end{equation}
%
\begin{equation}
\begin{split}
&{\cal V} = {\cal H}_{e} + {\cal H}_{p} = \gamma^2 \left( \stackrel{\circ}{\cal H}_{ex} + \stackrel{\circ}{\cal H}_{ey} \right) + \stackrel{\circ}{\cal H}_{ez} + \stackrel{\circ}{\cal H}_p ,\\
&{\cal T} = {\cal H}_{mx} + {\cal H}_{my} = \beta^2 \gamma^2 \left( \stackrel{\circ}{\cal H}_{ex} + \stackrel{\circ}{\cal H}_{ey} \right)
\end{split}
\end{equation}
and leads with $dz = \frac{d \stackrel{\circ}{z}}{\gamma}$ to the action
\begin{equation}
\begin{split}
S \; &= \; \int{d^4 r} \, {\cal L} \; = \; \int{d^4 r}  \left[ {\cal T} - {\cal V} \right] \; =\\
 \; &= \; - \int{\frac{d^4 \stackrel{\circ}{r}}{\gamma}} \left[ \gamma^2 (1 - \beta^2) \left( \stackrel{\circ}{\cal H}_{ex} + \stackrel{\circ}{\cal H}_{ey} \right) + \stackrel{\circ}{\cal H}_{ez} + \stackrel{\circ}{\cal H}_p \right] \; =\\
 \; &= \; - \int{d^4 \stackrel{\circ}{r}} \left[ \gamma (1 - \beta^2) \left( \stackrel{\circ}{\cal H}_{ex} + \stackrel{\circ}{\cal H}_{ey} \right) + \frac{1}{\gamma} \left( \stackrel{\circ}{\cal H}_{ez} + \stackrel{\circ}{\cal H}_p \right) \right] .
\end{split}
\end{equation}
From the variation of this action with respect to $\gamma$ we conclude
\begin{equation}
(1 - \beta^2) \left( \stackrel{\circ}{\cal H}_{ex} + \stackrel{\circ}{\cal H}_{ey} \right) \; = \; \frac{1}{\gamma^2} \left( \stackrel{\circ}{\cal H}_{ez} + \stackrel{\circ}{\cal H}_p \right) .
\end{equation}
Obviously, we get the solution
\begin{equation}
\gamma^2 \; = \; \frac{1}{1 - \beta^2} \quad \text{if} \quad  \stackrel{\circ}{\cal H}_{ex} \; = \; \stackrel{\circ}{\cal H}_{ey} \; = \; \stackrel{\circ}{\cal H}_{ez} \; = \; \stackrel{\circ}{\cal H}_{p} .
\end{equation}
The first two of these conditions are fulfilled due to the symmetry of the solution of the minimisation problem in the comoving frame. The last one is a consequence of the Derrick-Hobart necessity condition, as discussed at the end of Sect.~\ref{MonopoleSolutions}.

This discussion shows that it is possible to understand the relativistic behaviour of particles, e.g. Lorentz contraction and velocity dependence of mass, on a microsopic basis through the dynamics of the soliton field. Lorentz contraction turns out to be a consequence of the necessity to accelerate the soliton field. Additional electric curvature energy of $Q$ is needed for this acceleration process which can be transformed locally in magnetic (dynamic) energy. Both parts contribute to the relativistic energy of the moving particle
\begin{equation}
\begin{split}
m_v \; :&= \int{d^3 r} \left( {\cal T} + {\cal V} \right) \; 
= \gamma^2 \int{d^3 r} \left( \stackrel{\circ}{\cal H}_{ex} + \stackrel{\circ}{\cal H}_{ey} + \stackrel{\circ}{\cal H}_{ez} + \stackrel{\circ}{\cal H}_p \right) \; =\\
&= \; \gamma \int{d^3 \stackrel{\circ}{r}} \left( \stackrel{\circ}{\cal H}_{ex} + \stackrel{\circ}{\cal H}_{ey} + \stackrel{\circ}{\cal H}_{ez} + \stackrel{\circ}{\cal H}_p \right) \; = \; \gamma \, m_0
\end{split}
\end{equation}
with the well-known velocity dependence of the mass $m_v$ of the soliton.

\subsection{Four--dimensional formulation}\label{4dimForm}

In subsections \ref{SecElecMonop} - \ref{MovingSolitons} we gave a motivation for the choice of the field variables $Q$ and the Lagrangian of our model. To every point $x$ in space-time we assign an $SU(2)$-field, which can be identified with a unimodular quaternionic number. We want to take advantage of the topolgy of the parameter space $S^3$ of $Q$, the unit sphere in four dimensions. An easy way to work with these field variables is to use the fundamental representation of $SU(2)$ with Pauli-matrices $\vec{\sigma}$
\begin{equation}\label{Qreps}
\begin{split}
Q(x) =  e^{i \alpha(x) \vec{n}(x) \vec{\sigma}} = \; &q_0(x) + i \sigma_k q_k(x),\\
&q_0 = \cos \alpha, \; q_k = n_k \sin \alpha, \; \vec{n}^2 = 1 , \; 0 \le \alpha \le \frac{\pi}{2} .
\end{split}
\end{equation}
The Lagrangian
\begin{equation}\label{4dimL}
{\cal L} = {\cal L}_{dual} - {\cal H}_{p}
\end{equation}
consists of a conventional part
\begin{equation}\label{Ldual}
{\cal L}_{dual} =  - \frac{\alpha_f \hbar c}{4 \pi} \frac{1}{8} \; \text{Tr} R_{\mu \nu} R^{\mu \nu}
\end{equation}
and a potential term
\begin{equation}\label{PotentialTerm}
{\cal H}_{p} \; = \; \frac{\alpha_f \hbar c}{4 \pi} \, \Lambda(q_0(x))\; = \; \frac{\alpha_f \hbar c}{4 \pi r_0^4} \left( \frac{\text{Tr} Q(x)}{2} \right)^{2m} ,
\end{equation}
with Sommerfeld's finestructure constant $\alpha_f = \frac{e_0^2}{4 \pi \varepsilon_0 \hbar c }$.

The form of ${\cal L}_{dual}$ and of the curvature tensor
\begin{equation}\label{IntrinsicCurvature}
R_{\mu \nu} = - \partial_\mu \Gamma_\nu + \partial_\nu \Gamma_\mu + i \left[ \Gamma_\mu , \Gamma_\nu \right]
\end{equation}
is in full agreement with dual non-abelian $SU(2)$ gauge theory. In analogy to this theory we call therefore the internal space the colour space. The dual potential $C_\mu$ is related to the connection $\Gamma_\mu$ and the dual field strength tensor $\hspace{0.8mm}{^*}\hspace{-0.8mm}F_{\mu \nu}$ to the curvature $R_{\mu \nu}$
\begin{equation}\label{FinEandB}
C_\mu =  - \frac{e_0}{4 \pi \varepsilon_0} \Gamma_\mu , \quad
\hspace{0.8mm}{^*}\hspace{-0.8mm}F_{\mu \nu} = - \frac{e_0}{4 \pi \varepsilon_0 c} R_{\mu \nu} = 
\left( \begin{array}{cccc}
 0 & B_1 & B_2 & B_3 \\
 -B_1 & 0 & E_3/c & -E_2/c \\
 -B_2 & -E_3/c & 0 & E_1/c \\
 -B_3 & E_2/c & -E_1/c & 0 \\
\end{array} \right) .
\end{equation}
This agreement can also be nicely seen expressing the Lagrangian (\ref{4dimL}) in terms of the electric and magnetic field components $\vec{E}_k$ and $\vec{B}_k$
\begin{equation}
{\cal L} = \frac{1}{2} \left[ \frac{1}{\mu_0} ( \vec{B}_k )^2 - \varepsilon_0  ( \vec{E}_k )^2 \right] - {\cal H}_{p} .
\end{equation}
Furthermore, due to
\begin{equation}\label{ExtrinsicCurvature}
R_{\mu \nu} = - i [ \Gamma_\mu , \Gamma_\nu ] = ( \vec{\Gamma}_\mu \times \vec{\Gamma}_\nu ) \vec{T} = \vec{R}_{\mu \nu} \vec{T}
\end{equation}
${\cal L}_{dual}$ is identical to the Skyrme term in the Skyrme model. The agreement between expressions (\ref{IntrinsicCurvature}) and (\ref{ExtrinsicCurvature}) is caused by the Maurer-Cartan equation (\ref{MaurerCartan1}) and leads to an expression for the curvature tensor in terms of the rotor of the connection field
\begin{equation}\label{CurvAsRotor}
\vec{R}_{\mu \nu} =  - \frac{1}{2} ( \partial_\mu \vec{\Gamma}_\nu - \partial_\nu \vec{\Gamma}_\mu ) .
\end{equation}
The Maurer-Cartan equation (\ref{MaurerCartan1}) garantues that the soliton field $Q$ can be reconstructed from the connection
\begin{equation}\label{VecGamma}
\Gamma_\mu = \vec{\Gamma}_\mu \vec{T} , \quad \Gamma_{\mu k} =
 - i \, \text{Tr} \, ( \partial_\mu Q \, Q^\dagger \frac{\sigma_k}{2} ) .
\end{equation}

In the variables $\alpha$ and $\vec{n}$ of Eq.~(\ref{Qreps}) the connection reads
\begin{equation}\label{GammaMue}
\vec{\Gamma}_\mu =  \partial_\mu \alpha \, \vec{n} + \sin \alpha \cos \alpha \; \partial_\mu \vec{n} - \sin^2 \alpha \; \vec{n} \times \partial_\mu \vec{n} ,
\end{equation}
$\vec{n}$, $\partial_\mu \vec{n}$ and $\vec{n} \times \partial_\mu \vec{n}$ form an orthogonal Dreibein for $\vec{\Gamma}_\mu$. The square of the curvature in the Lagrangian (\ref{4dimL}) turns out to be a function of $q_0 = \cos \alpha$ and the derivatives $\partial_\mu q_0$ and $\partial_\mu \vec{n}$
\begin{equation}\label{SquareCurvature}
\begin{split}
\frac{1}{2} \; \text{Tr} R_{\mu \nu} R^{\mu \nu} &= \left( \vec{\Gamma}_\mu \times \vec{\Gamma}_\nu \right)^2 
= \vec{R}_{\mu \nu} \vec{R}^{\mu \nu} = 
\vec{\Gamma}_\mu^2 \vec{\Gamma}_\nu^2 - \left( \vec{\Gamma}_\mu \vec{\Gamma}_\nu \right)^2 =\\
&= \left( \partial_\mu q_0 \partial_\nu \vec{n} - \partial_\nu q_0 \partial_\mu \vec{n} \right)^2 + \left( 1 - q_0^2 \right)^2 \left( \partial_\mu \vec{n} \times \partial_\nu \vec{n} \right)^2
\end{split}
\end{equation}

${\cal L}_{dual}$ tends to decrease the curvature of the soliton field $Q$ by increasing the size of solitons. The potential term ${\cal H}_{p}$ on the other hand tries to diminish regions where $\text{Tr} Q$ is non-zero and to shrink solitons, therfore reducing the effective parameter space from $S^3$ to $S^2$ in space regions as large as possible.

We have discussed in Sect.~\ref{stabilization} that this competition between the curvature term ${\cal L}_{dual}$ and the potential term ${\cal H}_{p}$ leads to a variational problem for stable hedgehog solutions. The existence of non-trivial solutions we have shown in Sect.~\ref{MonopoleSolutions} by giving special examples. These solitonic solutions are characterised by a topological quantum number, the winding number $n_w$. We will demonstrate in Sec.~\ref{SectU1Gauge} that this model has a further topological quantum number, the electric charge. In Sec.~\ref{TopologicalCharge} we will show that the winding number agrees with the topological charge in its standard form. Since these topological quantum numbers are conserved independent of any dynamics they guarantee that the solitons survive scattering processes.

\section{Gauge transformations and $U(1)$-gauge theory}\label{SectU1Gauge}

With the Lagrangian (\ref{4dimL}) we succeeded to formulate a theory for topological solitons with a Coulomb field. Let us now investigate the relation of the $SU(2)$-field to the $U(1)$-field of charges in Maxwell's theory. According to the electro-weak model one could think of a $U(1) \times SU(2)$-structure of these degrees of freedom. But due to the cartesian product there are no special solitons which could be introduced by an additional $U(1)$-field and there is no simple possibility to transfer field strength from $SU(2)$ to $U(1)$-degrees of freedom. It turns out that this relation can be formulated in a much simpler way. At large distances from the center of a charged soliton the potential term ${\cal H}_{p}$ restricts the field degrees of freedom to an $S^2$-sphere, to the sphere $S^2_{int}$ with
\begin{equation}\label{asymptQ}
q_0 = 0 \quad \Longleftrightarrow \quad q_1^2 + q_2^2 + q_3^2 = 1 \quad \text{for} \quad r \rightarrow \infty .
\end{equation}
In this region the potential term can be neglected and the remaining degrees of freedom are the $\partial_\mu \vec{n}$-fields which appear in the expression (\ref{SquareCurvature}) for the square of the curvature. They dwell in the tangential spaces of the sphere $S^2_{int}$. On this sphere we can choose the local coordinate systems arbitrarily. Therefore, we arrive at a dual-$U(1)$-gauge theory. As we suggested in subsection~\ref{LocCoordSys} on the north-pole of this sphere we can identify the unit vectors in $\phi=0^\circ$ and $90^\circ$-directions with the real unit $1$ and the imaginary unit $i$. A parallel transport of this coordinate system along the meridians $\phi=const.$ inherits a local coordinate system to any position $(\theta, \phi)$ of $S^2_{int}$. The connection coefficients can easily be derived. We have obviously
\begin{equation}
\Gamma_\theta = 0 .
\end{equation}
A parallel transport of a vector along a parallel of latitude by the angle $d\phi$ leads to a (passive) rotation of the coordinate system by $-d\phi$. This value we have to decrease by the angle $d\phi \cos \theta$ for the rotation of the osculating great circle and get for the connection
\begin{equation}\label{U1connect}
\Gamma_\phi d\phi = - ( 1 - \cos \theta ) d\phi .
\end{equation}
If we take into account the spherical symmetry $(\theta=\vartheta, \phi=\varphi)$ of the soliton, multiply with the charge factor $- \frac{e_0}{4\pi \varepsilon_0}$ and divide by the corresponding distance $l_\varphi \ d \varphi = r \sin \vartheta \ d \varphi$ we get the dual vector potential
\begin{equation}\label{electricCpotential}
C_\vartheta = 0, \quad C_\varphi = - \frac{e_0}{4\pi \varepsilon_0} \frac{\Gamma_\phi}{r \sin \vartheta} = \frac{e_0}{4\pi \varepsilon_0} \frac{ 1 - \cos \vartheta }{r \sin \vartheta} .
\end{equation}
The dual vector-potential forms circles around the z-axis and is singular for $\vartheta = \pi$.  This potential is characteristic of a dual Dirac monopole, of an electric monopole of charge $- e_0$, with a singularity of the $U(1)$-potential along the negative z-axis. Its electric field strength is given by
\begin{equation}\label{DualEfield}
\vec{E} = - \vec{\nabla} \times \vec{C} = - \frac{e_0}{4 \pi \varepsilon_0}\frac{\vec{e}_r}{r^2}
\end{equation}
A formulation of electrodynamics in terms of dual potentials was given in Ref.\cite{BBZ94}.

In the following we will see that such a vector-potential  we can also get by a gauge-transformation of the connection field (\ref{sphericalVecPots}) which in the limit of large radial distances reads
\begin{equation}\label{U1potential}
\Gamma_r = 0 , \quad \Gamma_\vartheta = - \vec{T} \vec{e}_\varphi , \quad \Gamma_\varphi = \sin\vartheta \, \vec{T} \vec{e}_\vartheta ,\quad \text{for} \quad r \rightarrow \infty .
\end{equation}

We have seen in Eqs.~(\ref{tangentialVectors}) and (\ref{orthonormVectors}) that the vectors $\sigma_k^Q = \sigma_k Q$ form an orthonormal basis in the tangential space at $Q$. A gauge transformation is nothing else than a rotation of this basis in the tangential space with some $SU(2)$-matrix $\Omega$
\begin{equation}\label{rotBasis}
\sigma_k^{\prime Q} = \Omega^\dagger \sigma_k \, \Omega Q
\end{equation}
with
\begin{equation}
\begin{split}
\sigma_k^{\prime Q} \cdot \sigma_l^{\prime Q} &= \frac{1}{2} \text{Tr} \left\{ \Omega^\dagger \sigma_k \Omega Q Q^\dagger \Omega^\dagger \sigma_l \Omega \right\} = \frac{1}{2} \text{Tr} \left\{ \sigma_k \sigma_l \right\} = \delta_{kl} ,\\
\sigma_k^{\prime Q} \cdot Q &= \frac{1}{2} \text{Tr} \left\{ \Omega^\dagger \sigma_k \Omega Q Q^\dagger \right\} = \frac{1}{2} \text{Tr} \sigma_k = 0 . 
\end{split}
\end{equation}
The rotation of the basis vectors defines a matrix in the space of tangent vectors (in the adjoint representation of $SU(2)$) by
\begin{equation}\label{VectorRot}
\begin{split}
&\sigma_k^Q \quad \rightarrow \quad \sigma_k^{\prime Q} = \hat{\Omega}_{kl} \sigma_l^Q \qquad \text{with} \qquad \hat{\Omega} \hat{\Omega}^T = 1 ,\\
&\hat{\Omega}_{kl} = \sigma_k^{\prime Q} \cdot \sigma_l^{Q} = \frac{1}{2} \text{Tr} \left\{ \Omega^\dagger \sigma_k \Omega Q Q^\dagger \sigma_l \right\} = \frac{1}{2} \text{Tr} \left\{ \Omega^\dagger \sigma_k \Omega \sigma_l \right\} = \hat{\Omega}_{kl}^* . 
\end{split}
\end{equation}

We insert the transformation $\hat{\Omega}^\dagger \vec{\sigma}^{\prime Q} = \vec{\sigma}^Q$ in Eq.~(\ref{derivBasis}) and multiply from the left with $\hat{\Omega}$ and get
\begin{equation}\label{derivRotBasis}
\partial_s \vec{\sigma}^{\prime Q} = i \hat{\Omega} \left( \Gamma_s + i \partial_s \right) \hat{\Omega}^\dagger \vec{\sigma}^{\prime Q} + i \hat{\Omega} \vec{\Gamma}_s Q = i \Gamma_s^\prime \vec{\sigma}^{\prime Q} + i \vec{\gamma}_s^\prime Q .
\end{equation}
By comparing to Eq.~(\ref{derivBasis}) we read off that the $\Gamma$ in the tangential direction (the connection) transforms differently
\begin{equation}\label{PotgaugeTrafo}
\Gamma_s \rightarrow \Gamma_s^\prime = \hat{\Omega} \left( \Gamma_s + i \partial_s \right) \hat{\Omega}^\dagger
\end{equation}
from that in the radial direction
\begin{equation}\label{VectorTrafo}
\gamma_{sk} = \Gamma_{sk} \rightarrow \gamma_{sk}^\prime = \hat{\Omega}_{kl} \gamma_{sl} ,
\end{equation}
which transforms like a vector. Therefore, the Maurer-Cartan equation (\ref{MaurerCartan1}) is preserved for constant $\Omega$ only, i.e. for global gauge transformations.

A special gauge transformation is the transformation from the right adjoint basis $\sigma_k^Q := \sigma_k Q$ of Eq.~(\ref{rightadjoint}) to the left adjoint Basis $\sigma_k^{\prime Q} :=  Q \sigma_k$ by $\Omega = Q^\dagger$ in Eq.~(\ref{rotBasis}). The connection $\Gamma_s^\prime$ in the left adjoint basis we can read from Eq.~(\ref{PotgaugeTrafo}) using the fundamental representation $\Gamma_s = \vec{\Gamma}_s \frac{\vec{\sigma}}{2}$. With Eq.~(\ref{derivativeQ}) follows 
\begin{equation}
\partial_s Q = i \vec{\Gamma}_s \vec{\sigma}^Q = -i \vec{\Gamma}_s^\prime \vec{\sigma}^{\prime Q}
\end{equation}
and by forming the adjoint and by the transition $\alpha \rightarrow - \alpha$
\begin{equation}
\vec{\Gamma}_s^\prime (\alpha) = \vec{\Gamma}_s (-\alpha) .
\end{equation}

If we specify $\Omega=e^{i \vartheta \frac{\sigma_\varphi}{2}}$ or equivalently
\begin{equation}
\hat{\Omega} = e^{i \vartheta T_\varphi} = 1 + i \sin{\vartheta} \, T_\varphi + \left( cos \vartheta - 1 \right) T_\varphi^2 ,
\end{equation}
we get from (\ref{U1potential}) and (\ref{PotgaugeTrafo}) after some calculation
\begin{equation}
\Gamma_r^\prime = 0 , \quad \Gamma_\vartheta^\prime = 0 , \quad \Gamma_\varphi^\prime = \left( \cos \vartheta - 1 \right) T_3 .
\end{equation}

This result agrees with Eq.~(\ref{U1connect}). Therefore, our soliton at large distances behaves like a dual Dirac monopole. The Dirac string of the $U(1)$-formulation appears as a consequence of the the internal $SU(2)$-structure of the monopole. At large distances the soliton field of the monopole dwells at $S^2_{int}$
\begin{equation}
Q(x) \rightarrow i \vec{\sigma} \frac{\vec{r}}{r} =  i \vec{\sigma} \vec{n}, \quad \vec{\Gamma}_\mu(x)  \rightarrow \partial_\mu \vec{n} \times \vec{n} .
\end{equation}
The hedgehog field is completly regular, but turning all internal direction vectors $\vec{n}$ in z-direction leads to a singularity of the $U(1)$ field, to the Dirac string.

The $U(1)$ structure of solitons at large distances has interesting consequences. Solitons behave like electric charges. Positive and negative charges add like integer numbers. Solitons feel Coulomb- and Lorentz-forces.

At large distances from an ensemble of solitons and antisolitons we have to fulfil the condition (\ref{asymptQ}). This defines a mapping of the internal $S^2_{int}$ to the infinite far sphere  $S^2_{\infty}$ in coordinate space. It is well-known from topology that such non-singular mappings can be classified into homotopy classes. These homotopy classes form a group which in our case is isomorphic to the group of integers and can be written in the compact form \cite{St51}
\begin{equation}\label{ElectricCharge}
\pi_2(S^2)  =  Z .
\end{equation}
We conclude, due to their topological origin all electric charges are integer multiples of the unit charge $e_0$.

We get from Eq.~(\ref{GammaMue})
\begin{equation}\label{PreAbelianFlux}
\oint_{S(u,v)} du dv \, ( \vec{\Gamma}_u \times \vec{\Gamma}_v ) \, \vec{n} = \oint_{S(u,v)} du dv \, ( \partial_u \vec{n} \times \partial_v \vec{n} ) \, \vec{n} \sin^2 \alpha .
\end{equation}
This integral may take arbitrary values as long as $\sin^2 \alpha$ is arbitrary. Regions with $\cos \alpha \ne 0$ are suppressed by the potential energy (\ref{PotentialTerm}) in the Lagrangian. A region with $\cos \alpha \ne 0$ can in general decay in excitations with $\cos \alpha = 0$, i.e. in abelian excitations. This means that such a region radiates a conventional electro-magnetic field. This happens especially in the annihilation process of a soliton with an antisoliton. We conclude that regions with $\cos \alpha \ne 0$ survive in soliton cores only. If a surface $S(u,v)$ does not cross such soliton cores we get
\begin{equation}
\oint_{\substack{ S(u,v) \\ \cos \alpha = 0}} du dv \, R_{u v} =
\oint_{\substack{ S(u,v) \\ \cos \alpha = 0}} du dv \, ( \partial_u \vec{n} \times \partial_v \vec{n} ) \vec{T} .
\end{equation}
Since the cross-products of two tangential vectors of $S^2_{int}$ points in radial direction $\vec{n}$, the integral
\begin{equation}\label{AbelianFlux}
\oint_{\substack{ S(u,v) \\ \cos \alpha = 0}} du dv \, 
\frac{1}{2} \text{Tr} [R_{u v} T_n ] =
\oint_{\substack{ S(u,v) \\ \cos \alpha = 0}} du dv \, ( \partial_u \vec{n} \times \partial_v \vec{n} ) \vec{n} = 4 \pi Z
\end{equation}
adds up all the area elements $\partial_u \vec{n} \times \partial_v \vec{n} \, du dv $ of $S^2_{int}$ and counts the number $Z$ of its coverings, as we discussed above. Therefore, we define the electric charge by
\begin{equation}\label{ElecCharge}
- Z e_0 = - \frac{e_0}{8 \pi} \oint_{\substack{ S(u,v) \\ \cos \alpha = 0}} du dv \, \text{Tr} [R_{u v} T_n ] .
\end{equation}

By topological reasons there are no partial coverings of $S^2_{int}$ possible. Therefore, the soliton model can only describe integer charges, the electric charge is quantised. That is a difference to normal electrodynamics which can describe all sizes of charges. A $U(1)$ path integral over gauge degrees of freedom contains many unphysical states which do not fulfill Gau/{ss}'s law. To take into account physical states only one has to construct a projection operator which projects on states fulfilling Gau/{ss}'s law\cite{GPY81}.

Corresponding Eq.~(\ref{SquareCurvature}) in the Lagrangian outside of soliton cores there enter only derivatives of the soliton field
\begin{equation}
\frac{1}{2} \; \text{Tr} R_{\mu \nu} R^{\mu \nu} \rightarrow
\left( \partial_\mu \vec{n} \times \partial_\nu \vec{n} \right)^2 .
\end{equation}
Therefore, any change of the local coordinate system (\ref{derivRotBasis}) is allowed. This leads to the Lagrangian of a dual $U(1)$ gauge theory with integer multiples of the elementary charge only.

\section{Topological charge}\label{TopologicalCharge}

We can interpret the soliton field as a parametrisation of the internal space with external coordinates $x$, $y$, $z$ or $r$, $\vartheta$, $\varphi$. If we move in space by a line element $l_s \, ds$ we move in colour space by a line element $\vec{\Gamma}_s ds$. Obviously, the vectors $\vec{\Gamma}_r$, $\vec{\Gamma}_\vartheta$, $\vec{\Gamma}_\varphi$ form a (non orthonormal) coordinate system at the point $Q$ in internal space. The volume element in internal space corresponding to the volume element $l_r dr \, l_\vartheta d\vartheta \, l_\varphi d\varphi = dr \, r d\vartheta \, r \sin \vartheta d\varphi$ is therefore
\begin{equation}
\vec{\Gamma}_r (\vec{\Gamma}_\vartheta \times \vec{\Gamma}_\varphi) \; dr d\vartheta d\varphi .
\end{equation}
Integrating over this volume element we can define a topological charge ${\cal Q}$ counting the number of coverings of hemispheres  $S^3_{1/2}$
\begin{equation}
V(S^3_{1/2}) = \frac{1}{2} \int DQ = \frac{1}{2} \int_{S^2} d^2n \int_0^{\pi} d\alpha \sin^2 \alpha = \pi^2 
\end{equation}
in internal space $S^3$
\begin{equation}\label{TopolCharge}
{\cal Q} = \frac{1}{V(S^3_{1/2})} \int_0^\infty dr \int_0^\pi d\vartheta \int_0^{2 \pi} d\varphi \; \vec{\Gamma}_r (\vec{\Gamma}_\vartheta \times \vec{\Gamma}_\varphi) .
\end{equation}
This agrees with the usual definition of the topological charge given in ref. \cite{BPST75}. For the hedgehog-Ansatz (\ref{hedgehog}) and using Eq. (\ref{sphericalVecPots}) it reads
\begin{equation}
\begin{split}
{\cal Q} &= \frac{1}{\pi^2}  \int_0^\infty dr \int_0^\pi d\vartheta \int_0^{2 \pi} d\varphi \; \sin \vartheta \; \alpha^\prime (r) \sin^2 \alpha (r) = \\
&= \frac{4}{\pi} \int_{\alpha(0)}^{\alpha(\infty)} d \alpha \sin^2 \alpha = \frac{2}{\pi} \left[\alpha(\infty) - \alpha(0) \right] = n_w ,
\end{split}
\end{equation}
where we used the definition (\ref{WindingNumber}) for the winding number $n_w$.

Since the topological charge ${\cal Q}$ agrees with the winding number $n_w$ it characterises static solutions. Furthermore, as we will now discuss, any dynamic finite energy configuration can be characterised by the number ${\cal Q}$ of coverings of hemispheres $S^3_{1/2}$. As we discussed in Sec.~\ref{SectU1Gauge} due the potential term ${\cal H}_p$ in the Lagrangian (\ref{4dimL}) a dynamic field configuration has to converge for $r \rightarrow \infty$ to a $U(1)$ field configuration characterised by the electric charge $Ze_0$, the number $Z$ of coverings of $S^2_{int}$, of the sphere with $q_0 = \cos \alpha = 0$, as defined in Eq.~(\ref{ElectricCharge}). For finite energy configurations the space region with $q_0 \ne 0$ must be finite. Due to the curvature term ${\cal L}_{dual} = {\cal H}_m - {\cal H}_e$ in the Lagrangian neither $Z$ nor ${\cal Q}$ can be changed by any dynamics. $Z$ and ${\cal Q}$ are therefore invariant under any scattering process.

In the following we will prove the time independence of the topological charge ${\cal Q}$ in an analytical consideration. For simplicity we work in cartesian coordinates with $l_\mu = 1$. We introduce the topological current according Eq.~(\ref{TopolCharge})
\begin{equation}
k^\mu = \frac{1}{6 \pi^2} \epsilon^{\mu \nu \rho \sigma} \vec{\Gamma}_\nu (\vec{\Gamma}_\rho \times \vec{\Gamma}_\sigma),
\end{equation}
with the 4--dimensional antisymmetric tensor $\epsilon^{0123} = - \epsilon^{1230} = 1$. In terms of the time-component of the topological current the topological charge reads
\begin{equation}
{\cal Q}(t) = \int d^3r k^0(x) .
\end{equation}

With $\text{Tr} \left[ \sigma_a \sigma_b \sigma_c \right] = 2i \epsilon_{abc}$, Eq.(\ref{derivativeQ}) and $\partial_\mu Q Q^\dagger + Q \partial_\mu Q^\dagger = 0$ the topological current can be represented in an equivalent form
\begin{equation}\label{TheFamousTopologicalCurrent}
\begin{split}
\hspace{-0.2cm}&k^\mu = \frac{1}{6 \pi^2} \epsilon^{\mu \nu \rho \sigma} \vec{\Gamma}_\nu (\vec{\Gamma}_\rho \times \vec{\Gamma}_\sigma) = \frac{1}{12 \pi^2} \epsilon^{\mu \nu \rho \sigma} \text{Tr} [ (i \vec{\sigma} \vec{\Gamma}_\nu) (i \vec{\sigma} \vec{\Gamma}_\rho) (i \vec{\sigma} \vec{\Gamma}_\sigma) ] = \\
\hspace{-0.2cm}&= \frac{1}{12 \pi^2} \epsilon^{\mu \nu \rho \sigma} \text{Tr} [ \partial_\nu Q Q^\dagger \partial_\rho Q Q^\dagger \partial_\sigma Q Q^\dagger ] = \frac{1}{12 \pi^2} \epsilon^{\mu \nu \rho \sigma} \text{Tr} \left[ Q \partial_\nu Q^\dagger \partial_\rho Q \partial_\sigma Q^\dagger \right]  .
\end{split}
\end{equation}
which agrees completely with topological currents introduced in instanton physics \cite{BPST75} and the Skyrme model \cite{MRS93}.

The topological current (\ref{TheFamousTopologicalCurrent}) is obviously conserved
\begin{equation}\label{ConsTopCurr}
\begin{split}
\partial_\mu k^\mu &= \frac{1}{72 \pi^2} \epsilon^{\mu \nu \rho \sigma} \text{Tr} \left[ \partial_\mu Q \partial_\nu Q^\dagger \partial_\rho Q \partial_\sigma Q^\dagger \right] = \\
&= \frac{1}{72 \pi^2} \epsilon^{\mu \nu \rho \sigma} \text{Tr} \left[ \partial_\mu Q Q^\dagger \partial_\nu Q Q^\dagger  \partial_\rho Q Q^\dagger \partial_\sigma Q Q^\dagger \right] = 0 .
\end{split}
\end{equation}
The r.h.s of this equation is zero due to the symmetry properties of the trace and the tensor $\epsilon^{\mu \nu \rho \sigma}$. This entails the conservation of the topological charge
\begin{equation}\label{ConsTopCharge}
\frac{d{\cal Q}(t)}{d t} = \int d^3r\,\frac{\partial}{\partial
t}k^0(x) = c \int d^3r\,\partial_\mu k^\mu(x) = 0 ,
\end{equation} 
if no soliton is allowed to leave the system. This implies that the stabilizing potential energy has great influence on the conservation of the topological charge. In fact, due to the stabilizing potential ${\cal H}_p$ the solution of the classical equations of motion acquires the shape of a localised soliton with the spatial current $\vec{k}$, the spatial part of a topological current, vanishing at $r \rightarrow \infty$ steeper than $1/r^2$. This has allowed to use 
\begin{equation}
\int d^3r \; div \; \vec{k}(x) = \oint_{S^2_\infty} d\vec{S} \; \vec{k}(x) = 0
\end{equation} 
for the derivation of Eq.~(\ref{ConsTopCharge}).

There is an interesting consequence of the conservation of the topological charge. We can conclude that $F \hspace{0.5mm}{^*}\hspace{-0.8mm}F$ vanishes identically: This follows with Eq.(\ref{derivativeQ}) from the conservation (\ref{ConsTopCurr})
\begin{equation}
\begin{split}
0 &= \epsilon^{\mu \nu \rho \sigma} \text{Tr} \left[ \partial_\mu Q Q^\dagger \partial_\nu Q Q^\dagger  \partial_\rho Q Q^\dagger \partial_\sigma Q Q^\dagger \right] =\\
&= \epsilon^{\mu \nu \rho \sigma} \text{Tr} [ (i \vec{\sigma} \vec{\Gamma}_\mu) (i \vec{\sigma} \vec{\Gamma}_\nu) (i \vec{\sigma} \vec{\Gamma}_\rho) (i \vec{\sigma} \vec{\Gamma}_\sigma) ] =\\
&= - 2 \, \epsilon^{\mu \nu \rho \sigma} (\vec{\Gamma}_\mu \times \vec{\Gamma}_\nu) (\vec{\Gamma}_\rho \times \vec{\Gamma}_\sigma) = - \epsilon^{\mu \nu \rho \sigma} \text{Tr} [ R_{\mu \nu} R_{\rho \sigma} ] \propto \text{Tr} [ F^{\mu \nu} \hspace{0.5mm}{^*}\hspace{-0.8mm}F_{\mu \nu} ] ,
\end{split}
\end{equation} 
where
\begin{equation}\label{dualTrafo}
F^{\mu \nu} = - \frac{1}{2} \epsilon^{\mu \nu \rho \sigma} \hspace{0.8mm}{^*}\hspace{-0.8mm}F_{\rho \sigma} = 
\left( \begin{array}{cccc}
 0 & -E_1/c & -E_2/c & -E_3/c \\
 E_1/c & 0 & -B_3 & B_2 \\
 E_2/c & B_3 & 0 & -B_1 \\
 E_3/c & -B_2 & B_1 & 0 \\
\end{array} \right) .
\end{equation}
This vanishing of $F \hspace{0.5mm}{^*}\hspace{-0.8mm}F$ is identical to the statements that electric field strengh $\vec{E}$ and magnetic field strengh $\vec{B}$ are everywhere in space-time orthogonal to each other and that there is no anomaly.

Another interesting property which follows from the identification of the field-strenght tensor with the curvature of the soliton field is the conservation of electric flux. This is the dual Bianchi identity. It can be derived from Eqs.~(\ref{IntrinsicCurvature}) and (\ref{ExtrinsicCurvature})
\begin{equation}
\epsilon^{\mu \nu \rho \sigma} \partial_\nu R_{\rho \sigma} = \epsilon^{\mu \nu \rho \sigma} \partial_\nu (\vec{\Gamma}_\rho \times \vec{\Gamma}_\sigma) \vec{T} = - \frac{1}{2} \epsilon^{\mu \nu \rho \sigma} \partial_\nu (\partial_\rho \Gamma_\sigma - \partial_\sigma \Gamma_\rho) = 0
\end{equation}
and the symmetry of double derivatives. Therefore, the non-abelian electric current
\begin{equation}\label{ColElCurr}
\vec{j}_{el}^\nu = \partial_\mu \vec{F}^{\mu \nu} = 0
\end{equation}
vanishes identically. From the vanishing of the non-abelian electric charge follows that there are no sources for a non-abelian electric field and the non-abelian electric flux-lines are closed
\begin{equation}
\partial_i \vec{E}_i = \partial_i \epsilon_{ijk} \partial_j \vec{C}_k = 0 .
\end{equation}

This does not mean that the abelian electric flux through such closed surfaces is vanishing, as we discussed at the end of Sect.~\ref{SectU1Gauge}.

\section{Equations of motion}

Another interesting insight in the model we get from the equation of motion. In Sect.~\ref{stabilization} we discussed the variational problem for the hedgehog ansatz (\ref{hedgehog}) which leads to the non-linear differential equation (\ref{nlDE}). Now we are going to investigate the general variatonal problem of our model.

We vary the soliton field $Q(x)$ by $\delta Q$ in the tangential plane at $Q$
\begin{equation}\label{deltaQ}
\delta Q  =  i \, \delta \vec{\Gamma} \, \vec{\sigma} \, Q
\end{equation}
and get the quantities
\begin{equation}\label{deltaVecGamma}
\delta \vec{\Gamma} = - i \, \text{Tr} \, ( \delta  Q \, Q^\dagger \frac{\vec{\sigma}}{2} ) , \qquad 
\delta q_0 = \frac{1}{2} \text{Tr} \, \delta  Q = - \, \delta \vec{\Gamma} \, \vec{q}
\end{equation}
and from (\ref{VecGamma}) follows the variation of the connection field
\begin{equation}\label{deltaVecGammaMue}
\begin{split}
&\delta \vec{\Gamma}_\mu =
\text{Tr} \, ( \partial_\mu \delta Q \, Q^\dagger \frac{\vec{\sigma}}{2i} ) + \text{Tr} \, ( \partial_\mu Q \, \delta Q^\dagger \frac{\vec{\sigma}}{2i} ) = \\
&= \text{Tr} \, ( [ \delta Q \, Q^\dagger , \partial_\mu Q \, Q^\dagger ] \frac{\vec{\sigma}}{2i} ) - \text{Tr} \, ( \delta Q \, Q^\dagger \frac{\vec{\sigma}}{2i} ) \, \partial_\mu = - \, \delta \vec{\Gamma} \, \partial_\mu - 2 \, \delta \vec{\Gamma} \times \vec{\Gamma}_\mu .
\end{split}
\end{equation}
The variation of the curvature term of the Lagrangian (\ref{4dimL}) reads
\begin{equation}\label{deltaCurv}
\delta [ \frac{1}{4} \vec{R}_{\mu \nu} \vec{R}^{\mu \nu} ] =
\delta \vec{\Gamma}_\mu [ \vec{\Gamma}_\nu \times \vec{R}^{\mu \nu} ] = - \delta \vec{\Gamma} \, \partial_\mu [ \vec{\Gamma}_\nu \times \vec{R}^{\mu \nu} ] ,
\end{equation}
and that of the potential term (\ref{PotentialTerm})
\begin{equation}\label{deltaPot}
\delta \Lambda = \frac{d \Lambda}{d q_0} \, \delta q_0 = - \delta \vec{\Gamma} \, \vec{q} \, \frac{d \Lambda}{d q_0} .
\end{equation}
Since the variation $\delta \vec{\Gamma}$ is arbitrary the equation of motion follows
\begin{equation}\label{EqMotionNatural}
\partial_\mu [ \vec{\Gamma}_\nu \times \vec{R}^{\mu \nu} ] \, + \, \vec{q} \, \frac{d \Lambda}{d q_0} \, = 0 .
\end{equation}
From this equation we read that there is a current
\begin{equation}\label{MagCurrent}
\vec{J}^\mu = \vec{\Gamma}_\nu \times \vec{R}^{\mu \nu} 
\end{equation}
which is conserved in regions of vanishing potential energy ${\cal H}_{p}$, i.e. outside of soliton cores. Soliton cores are a source of this current
\begin{equation}\label{PartConsCurrent}
\partial_\mu \vec{J}^\mu = - \vec{q} \, \frac{d \Lambda}{d q_0} .
\end{equation}
In a static soliton only the space components of $\vec{\Gamma_\mu}$ and $\vec{R}^{\mu \nu}$ are non-vanishing. Therefore, the time-component of $\vec{J}$ vanishes identically. This leads to the interpretation that $\vec{J}$ is a stationary magnetic current inside the soliton and the origin of its static electric field.

In physical units this magnetic current reads
\begin{equation}\label{SIMagCurrent}
\vec{j}^\mu = \frac{e_0 c}{2 \pi} \vec{J}^\mu .
\end{equation}
With his help and the relation (\ref{FinEandB}) between $\hspace{0.5mm}{^*}\hspace{-0.8mm}F_{\mu \nu}$ and $\vec{R}^{\mu \nu}$ we get a very simple form for the Lagrange density (\ref{4dimL})
\begin{equation}
{\cal L} = - \frac{1}{8} \vec{C}_\mu \vec{j}^\mu  - {\cal H}_{p} .
\end{equation}

There is another nice interpretation of the equation of motion (\ref{EqMotionNatural}). With expression (\ref{CurvAsRotor}) for the curvature and using $\vec{R}_{\mu \nu} \times \vec{R}^{\mu \nu} = 0$ we conclude
\begin{equation}\label{OrthMagCurr}
\vec{\Gamma}_\nu \times \partial_\mu \vec{R}^{\mu \nu}\, + \, \vec{q} \, \frac{d \Lambda}{d q_0} \, = 0 .
\end{equation}
This equation suggests to introduce the obviously conserved non-abelian magnetic current
\begin{equation}\label{ConsMagCurr}
\vec{j}_{mag}^\nu = \partial_\mu \hspace{0.5mm}{^*}\hspace{-0.8mm}\vec{F}^{\mu \nu} , \qquad \partial_\nu \vec{j}_{mag}^\nu = 0.
\end{equation}
According to the equation of motion in regions of vanishing potential energy ${\cal H}_{p}$ the non-abelian magnetic current $\vec{j}_{mag}^\nu$ and the dual vector potential $\vec{C}^\nu$ have the same direction in colour space. Inside of solitons there is also a transversal colour component
\begin{equation}
[ j_{mag}^\nu , C_\nu ] = i \, \frac{d {\cal H}_{p}}{d q_0} \, \frac{q}{\varepsilon_0} \qquad \text{with} \qquad q = \vec{q} \vec{T}.
\end{equation}

\section{Energy--momentum tensor}\label{EnergyMomentumTensor}

In this Section we derive the explicit expression for the energy--momentum tensor $\Theta_{\mu\nu}(x)$. With (\ref{SquareCurvature}) and (\ref{PotentialTerm}) we write the Lagrangian (\ref{4dimL}) in the form
\begin{equation}
{\cal L} =  - \frac{\alpha_f \hbar c}{4 \pi} \left\{ \frac{1}{4} \left( \vec{\Gamma}_\rho \times \vec{\Gamma}_\sigma \right)^2  + \frac{1}{r_0^4} q_0^{2m} \right\} .
\end{equation}
Regarding $\alpha(x)$, $\partial_\mu \alpha(x)$ and $\partial_\mu \vec{n}(x)$ as the field variables we use (\ref{GammaMue}) and
\begin{equation}
\begin{split}
\frac{\partial \vec{\Gamma}_\rho}{\partial (\partial_\mu \alpha)} &\partial_\nu \alpha +
\frac{\partial \vec{\Gamma}_\rho}{\partial (\partial_\mu n_k)} \partial_\nu n_k = \\
&= ( \partial_\nu \alpha \, \vec{n} + \sin \alpha \cos \alpha \; \partial_\nu \vec{n} - \sin^2 \alpha \; \vec{n} \times \partial_\nu \vec{n} ) \delta^\mu_\rho =
\vec{\Gamma}_\nu \, \delta^\mu_\rho
\end{split}
\end{equation}
for the determination of the energy--momentum tensor
\begin{equation}\label{DefEMT}
\Theta^\mu_{\;\nu}(x) = 
\frac{\partial {\cal L}(x)}{\partial (\partial_\mu \alpha)} \partial_\nu \alpha + 
\frac{\partial {\cal L}(x)}{\partial (\partial_\mu n_k)} \partial_\nu n_k - 
{\cal L}(x)\,\delta^\mu_\nu
\end{equation}
and the metric tensor $\eta_{\mu\nu} = \text{diag}(1,-1,-1,-1)$. According to the symmetry of the expression we get
\begin{equation}\label{ExpressionEMT}
\Theta^\mu_{\;\nu} =
- \frac{\alpha_f \hbar c}{4 \pi}
\left\{ \left( \vec{\Gamma}_\nu \times \vec{\Gamma}_\sigma \right)\left( \vec{\Gamma}^\mu \times \vec{\Gamma}^\sigma \right) \right\} - 
{\cal L}(x)\,\delta^\mu_\nu .
\end{equation}
We would like to emphasize that in Classical Electrodynamics \cite{Ja75} the energy--momentum tensor should be specially symmetrised.

Now we consider the various components of the energy--momentum tensor $\Theta^\mu_{\;\nu}$ and express them in terms of electric and magnetic fields (\ref{FinEandB}). $\Theta^0_{\;0}$ is the Hamilton density ${\cal H}$ in full agreement with Sec.~\ref{Formulation}
\begin{equation}\label{Theta00}
\Theta^0_{\;0} = {\cal H} = \frac{1}{2} \left[ \frac{1}{\mu_0} ( \vec{B}_k )^2 + \varepsilon_0 ( \vec{E}_k )^2 \right] + {\cal H}_{p} .
\end{equation}
The space-time components read in complete analogy to the Poynting vector of electrodynamics
\begin{equation}\label{Theta0i}
\Theta^0_{\;i} = c \, \varepsilon_0 \, \epsilon_{ijk} \vec{B}_j \vec{E}_k .
\end{equation}
For the $ij$-components we get
\begin{equation}\label{Thetaij}
\Theta^i_{\;j} =\varepsilon_0 \vec{E}_i \vec{E}_j + \frac{1}{\mu_0} \vec{B}_i \vec{B}_j + [ {\cal H}_p - \frac{\varepsilon_0}{2} \vec{E}_k^2 - \frac{1}{2 \mu_0} \vec{B}_k^2 ] \, \delta^i_j .
\end{equation}

These equations generalize the expressions for the energy-momentum tensor of Classical Electrodynamics \cite{Ja75} to the inclusion of a stabilizing energy ${\cal H}_p(x)$ and internal $SU(2)$ degrees of freedom.

\section{Relativistic mechanics of a topological electron}\label{RelMech}

From the energy--momentum tensor $\Theta^\mu_{\;\nu}(x)$ of Sect.~\ref{EnergyMomentumTensor} we are now going to derive the 4--momentum of a topological electron. Like in Sect.~\ref{MovingSolitons} we treat a soliton moving in the laboratory system with constant velocity $\vec{\beta}=\vec{v}/c$. The three-dimensional space in the comoving frame is characterised by its normal
\begin{equation}
\stackrel{\circ}{\eta} ^\nu(x)=(1,0,0,0)
\end{equation}
in four--dimensional  Minkowski space. In the laboratory sytem $\eta^{\nu}(x)$ has the components
\begin{equation}\label{eta}
\eta^{\nu}(x) = (\gamma, \gamma\,\vec{\beta}\,).
\end{equation}
With the invariant 3--volume element $d^3\stackrel{\circ}{r}$ in the comoving frame we define the 4--vector 
\begin{equation}\label{d3r}
d\sigma^{\nu}(x) = \eta^{\nu}(x)\,d^3\!\!\stackrel{\circ}{r}
\end{equation}
and the 4--momentum
\begin{equation}\label{4Momentum}
P^{\mu} = \frac{1}{c} \int \Theta^\mu_{\;\nu}\,\eta^{\nu}(x)\,d^3\!\!\stackrel{\circ}{r}.
\end{equation}
Inserting Eqs.~(\ref{Theta00}-\ref{Thetaij}) we get the components of the momentum 4--vector in terms of electric and magnetic fields
\begin{equation}\label{Comp4Momentum}
\begin{split}
P^0 = \frac{\gamma}{c} \int \,d^3\!\!\stackrel{\circ}{r}
&\{ \frac{  \varepsilon_0}{2} ( \vec{E}_k )^2 + \frac{1}{2 \mu_0} ( \vec{B}_k )^2 + {\cal H}_{p} + c \varepsilon_0 \, \epsilon_{ijk} \beta^i \, \vec{B}_j \vec{E}_k \} ,\\
P^i = \frac{\gamma}{c} \int \,d^3\!\!\stackrel{\circ}{r}
&\{ - c \, \varepsilon_0 \, \epsilon_{ijk} \vec{B}_j \vec{E}_k +
\beta^j \varepsilon_0 \vec{E}_i \vec{E}_j + \beta^j \frac{1}{\mu_0} \vec{B}_i \vec{B}_j +\\
&+ \beta^i \, [ {\cal H}_p - \frac{\varepsilon_0}{2} \vec{E}_k^2 - \frac{1}{2 \mu_0} \vec{B}_k^2 ] \}.
\end{split}
\end{equation}
For  configurations static in the comoving reference frame we derived in Eq.~(\ref{BeqBetaXE}) a relation between the magnetic and the electric field which we can write in more general form%
\begin{eqnarray}\label{BField}
\vec{B}_i(x) = \epsilon_{ijk} \beta^j \vec{E}_k(x).
\end{eqnarray}
Inserting this equation in Eq.~(\ref{Comp4Momentum}) we get
\begin{equation}\label{Comp4Momentum2}
\begin{split}
P^0 &= - \frac{\gamma}{c} \int \,d^3\!\!\stackrel{\circ}{r}
[ \frac{1}{2 \mu_0} ( \vec{B}_k )^2 - \frac{\varepsilon_0}{2} ( \vec{E}_k )^2 - {\cal H}_p ] 
= - \frac{\gamma}{c} \int \,d^3\!\!\stackrel{\circ}{r} {\cal L}
,\\
P^i &= \beta^i \, \frac{\gamma}{c} \int \,d^3\!\!\stackrel{\circ}{r}
[ \frac{1}{2 \mu_0} ( \vec{B}_k )^2 - \frac{\varepsilon_0}{2} ( \vec{E}_k )^2 - {\cal H}_p ] 
= - \beta^i \, \frac{\gamma}{c} \int \,d^3\!\!\stackrel{\circ}{r} {\cal L} .
\end{split}
\end{equation}
Since the Lagrange density is a Lorentz scalar we can evaluate it in the comoving coordinate system. There it is equal to the negative value of the Hamiltonian density
\begin{equation}\label{Comp4Momentum3}
\begin{split}
P^0 &= - \frac{\gamma}{c} \int \,d^3\!\!\stackrel{\circ}{r} \stackrel{\circ}{\cal L}
= \frac{\gamma}{c} \int \,d^3\!\!\stackrel{\circ}{r} \stackrel{\circ}{\cal H} = \gamma \, m_0 c ,\\
P^i &= - \beta^i \, \frac{\gamma}{c} \int \,d^3\!\!\stackrel{\circ}{r} {\cal L}
= \beta^i \, \frac{\gamma}{c} \int \,d^3\!\!\stackrel{\circ}{r} \stackrel{\circ}{\cal H} = \beta^i \, \gamma \, m_0 c .
\end{split}
\end{equation}
These formulae confirm that a topological electron moving with velocity $c\,\vec{\beta}$ respects the relation between the energy $P^0 = \gamma\,m_0 c^2$ and the 3--momentum $P^i = \beta^i \, \gamma \, m_0 c$ of a classical particle
\begin{equation}
P_{\mu} P^{\mu} = m^2_0 c^2 .
\end{equation}

This shows that solitons, topological stable objects, behave like classical particles and can scatter at each other. In the next section we are going to show that the interaction of these solitons with external electric and magnetic fields follows Newton's equation of motion. Solitons experience Coulomb and Lorentz forces.

\section{Interaction with external fields}

The forces acting at solitons follow from the divergence of the energy--momentum tensor (\ref{ExpressionEMT}) which is the vector $f$ of the force density
\begin{equation}
\begin{split}
f_{\nu} = \, \partial_\mu \Theta^\mu_{\;\nu} = &- \frac{\alpha_f \hbar c}{4 \pi}
 \{ \partial_\mu \vec{\Gamma}_\nu [ \vec{\Gamma}_\sigma \times ( \vec{\Gamma}^\mu \times \vec{\Gamma}^\sigma ) ] + \\
&+ \vec{\Gamma}_\nu \partial_\mu [ \vec{\Gamma}_\sigma \times ( \vec{\Gamma}^\mu \times \vec{\Gamma}^\sigma ) ] - \partial_\nu \vec{\Gamma}_\rho  [ \vec{\Gamma}_\sigma \times ( \vec{\Gamma}^\rho \times \vec{\Gamma}^\sigma ) ] - \partial_\nu \Lambda \}
.
\end{split}
\end{equation}
After inserting the definition (\ref{MagCurrent}) of the current $\vec{J}^\mu$ we get
\begin{equation}
f_{\nu} = \, - \frac{\alpha_f \hbar c}{4 \pi}
 \{ \partial_\mu \vec{\Gamma}_\nu \vec{J}^\mu
 + \vec{\Gamma}_\nu \partial_\mu \vec{J}^\mu
 - \partial_\nu \vec{\Gamma}_\mu  \vec{J}^\mu - \partial_\nu \Lambda \}
\end{equation}
and with the curvature tensor in terms of a rotor (\ref{CurvAsRotor})
\begin{equation}\label{ForceDensity}
f_{\nu} = \, - \frac{\alpha_f \hbar c}{4 \pi}
 \{ 2 \vec{R}_{\nu\mu} \vec{J}^\mu + \vec{\Gamma}_\nu \partial_\mu \vec{J}^\mu - \partial_\nu \Lambda \} .
\end{equation}
We are now going to show that this expression vanishes for a closed system. Its first term is zero
\begin{equation}\label{RJINteraction}
\vec{R}_{\nu\mu} \vec{J}^\mu = ( \vec{\Gamma}_\nu \times \vec{\Gamma}_\mu ) [ \vec{\Gamma}_\sigma \times \vec{R}^{\mu\sigma} ] = ( \vec{\Gamma}_\nu \vec{\Gamma}_\sigma ) ( \vec{\Gamma}_\mu \vec{R}^{\mu\sigma} ) - ( \vec{\Gamma}_\mu \vec{\Gamma}_\sigma ) ( \vec{\Gamma}_\nu \vec{R}^{\mu\sigma} ) = 0 
\end{equation}
due to the orthogonality of $\vec{\Gamma}_\mu$ and $\vec{R}^{\mu\sigma}$ and due to the contraction of the antisymmetric tensor $\vec{R}^{\mu\sigma}$ with the symmetric tensor $\vec{\Gamma}_\mu \vec{\Gamma}_\sigma$. The second term $\vec{\Gamma}_\nu \partial_\mu \vec{J}^\mu$ of Eq.~(\ref{ForceDensity}) agrees with the third term $\partial_\nu \Lambda$ due to the equation of motion (\ref{PartConsCurrent})
\begin{equation}\label{GammaDJInt}
\vec{\Gamma}_\nu \partial_\mu \vec{J}^\mu = - \vec{\Gamma}_\nu \vec{q} \, \frac{d \Lambda}{d q_0} = - \partial_\nu \alpha \sin \alpha \, \frac{d \Lambda}{d q_0} = \partial_\nu q_0 \, \frac{d \Lambda}{d q_0} = \partial_\nu \Lambda .
\end{equation}
We conclude that the force density for a closed system vanishes everywhere,
\begin{equation}
f_{\nu} = 0 .
\end{equation}

In order to describe the influence of external fields we have to decompose them from internal fields which have their origin in the internal current $\vec{J}^\mu$. This is only approximately possible. Since $\vec{J}^\mu$ is not related to an external field $\vec{R}_{\nu\mu}^{ext}$ the term (\ref{RJINteraction}) is not vanishing and leads to an external force%
\begin{equation}\label{ExtCurvForce}
F_{\nu}^{ext} =  \, - \frac{\alpha_f \hbar c}{2 \pi} \int d^3r \vec{R}_{\nu\mu}^{ext} \vec{J}^\mu = \int \hspace{0.5mm}{^*}\hspace{-0.8mm}\vec{F}_{\nu\mu}^{ext} d^3r \vec{j}^\mu .
\end{equation}
This external force describes the interaction of a charge distribution with external electromagnetic fields and has a form which resembles the well-known Coulomb and Lorentz forces. But the realisation of these forces is dual to that in electrodynamics. As one can immediately see from Eqs.~(\ref{MagCurrent}) and (\ref{SIMagCurrent}) a static soliton is characterised by a vanishing time component $\vec{j}^0$ and non-vanishing space-components $\vec{j}^i$. This spatial current interacts with the external electric field which is represented by a curvature of the soliton field created by an external source. Due to the Lorentz invariance of the formulation it follows that moving solitons feel a Lorentz force of a strength proportional to their velocity. This force appears also in dual form as an interaction of the time component of the soliton current with the time component of the curvature field.

There may be a further interaction due to a possible violation of Eq.~(\ref{GammaDJInt}) by external fields. This interaction is due to a non-vanishing external connection field $\vec{\Gamma}_\nu$ at non-vanishing $\partial_\mu \vec{J}^\mu$.

In the limit of distances large compared to the fundamental length $r_0$ the soliton field degenerates to $Q=i \vec{\sigma} \vec{n}$ and solitons can be described by world--lines $X^\mu(\tau)$ and the action of a free solitonic particle of mass $m_0$
\begin{equation}
S_{fp} = - m_0 c^2 \int d\tau \int d^4x \, \delta^4(x-X(\tau)) \sqrt{\frac{dX_\mu}{d\tau}\frac{dX^\mu}{d\tau}} .
\end{equation}
In this limit it is possible to show that electro--magnetic forces are a consequence of topology \cite{HT93}. Let us follow these arguments.

We define an abelian field strength $f_{\mu \nu}$ by
\begin{equation}\label{AbelianFS}
{\hspace{1mm}^*}\hspace{-1mm}f_{\mu \nu}(x) = - \frac{e_0}{4 \pi \epsilon_0 c} [ \partial_\mu \vec{n}(x) \times \partial_\nu \vec{n}(x) ] \vec{n}(x)
\end{equation}
and get from Eq.~(\ref{Ldual}) the action of a free abelian field
\begin{equation}
S_{ff} = - \frac{1}{4 \mu_0} \int d^4x  {\hspace{1mm}^*}\hspace{-1mm}f_{\mu \nu}(x) \hspace{1mm}{^*}\hspace{-1mm}f^{\mu \nu}(x) = \frac{1}{4 \mu_0} \int d^4x \, f_{\mu \nu}(x) f^{\mu \nu}(x) .
\end{equation}
From Eq.~(\ref{AbelianFlux}) we obtain the flux condition
\begin{equation}
\oint_{S(u,v)} \hspace{-6mm} du dv \hspace{0.8mm}{^*}\hspace{-1mm}f_{u v}(x) = - \frac{e_0}{\epsilon_0} Z(S)
\end{equation}
where $Z$ is the charge inside of the closed surface $S(u,v)$ parametrised by $u$ and $v$. Since this equation holds for any volume it holds also differentially
\begin{equation}\label{FSInteraction}
\partial_\mu f^{\mu \nu}(x) = - \frac{e_0}{\epsilon_0 c} \int d\tau \frac{X^\nu(\tau)}{d \tau} \delta^4 (x-X(\tau)) .
\end{equation}
In a variational procedure we can add to the action of the free field and of the free soliton the constraint (\ref{FSInteraction}) with a Lagrange multiplier $a_\mu (x)$ and get the action
\begin{equation}\label{AbAction}
\begin{split}
S = & \int d^4x \{ \frac{1}{\mu_0} [ \frac{1}{4} f_{\mu \nu}(x) f^{\mu \nu}(x) + a_\nu (x) \, \partial_\mu f^{\mu \nu}(x) ]  + \\
&+ \int d\tau [ e_0 c \, a_\nu (x) \frac{dX^\nu(\tau)}{d\tau} \, - \, m_0 c^2 \sqrt{\frac{dX_\mu}{d\tau}\frac{dX^\mu}{d\tau}} ] \delta^4(x-X(\tau)) \} .
\end{split}
\end{equation}
Extremising $S$ with respect to $X^\mu(\tau)$ leads to
\begin{equation}\label{NewEqSol}
m_0 \, \frac{d^2 X^\mu(\tau)}{d\tau^2} = e_0 c [ \partial^\mu a^\nu(X(\tau)) - \partial^\nu a^\mu(X(\tau)) ] \frac{dX_\nu}{d\tau} .
\end{equation}
By a variation with respect to $\partial_\mu \vec{n}$ one obtains the relation between $f_{\mu \nu}(x)$ and $a_\mu(x)$
\begin{equation}
f_{\mu \nu}(x) = \partial_\mu a_\nu - \partial_\nu a_\mu
\end{equation}
which shows that the Lagrange multiplier appears as the gauge potential of an abelian field and the action (\ref{AbAction}) is that of an abelian theory. The phase space of this abelian theory is restricted to physical states \cite{GPY81} with charges restricted to integer multiples of the elementary charge $e_0$, as was discussed in Sect.~\ref{SectU1Gauge}.

Eq.~(\ref{NewEqSol}) is obviously Newtons equation of motion of a particle in an ``external'' field $f^{\mu \nu}$
\begin{equation}\label{NewEqSol2}
m_0 \, \frac{d^2 X^\mu(\tau)}{d\tau^2} = e_0 c f^{\mu \nu}(X(\tau)) \frac{dX_\nu}{d\tau}
\end{equation}
with $f^{\mu \nu}$ according to Eq.~(\ref{AbelianFS})
\begin{equation}
f^{\mu \nu}(x) = \frac{e_0}{4 \pi \epsilon_0 c} \frac{1}{2} \epsilon^{\mu \nu \rho \sigma} [ \partial_\rho \vec{n}(x) \times \partial_\sigma\vec{n}(x) ] \vec{n}(x) .
\end{equation}
Solitons feel Coulomb- and Lorentz forces. Their interaction with the electro--magnetic field is only a consequence of topology.

\section{Types of stable soliton configurations}\label{Types}

The topological charge ${\cal Q}$ can be used to classify stable $SU(2)$ soliton configurations which are solutions of the variational problem (\ref{nlDE}). In Sect.~\ref{MonopoleSolutions} we discussed stable field configurations of hedgehog type (\ref{hedgehog}). For these configurations the colour direction $\vec{n}$ agrees with the unit vector in space $\vec{r}/r$. Stable soliton configurations exist for ${\cal Q} = 1, \,  3, \, 5, \cdots$. As defined in Eq.~(\ref{ElecCharge}) these configurations are also characterised by the electric charge $-e_0$.

By a global rotation of the local coordinate system the connection $\Gamma$ transforms like a vector (\ref{VectorTrafo}). Therefore, electric and topological charge are invariants under such global gauge transformations.

\begin{figure}
\centerline{\scalebox{0.5}{\includegraphics{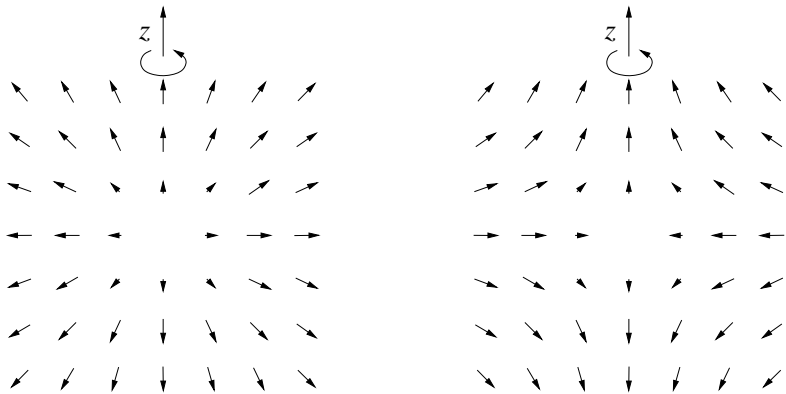}}}
\caption{Schematic diagrams of the imaginary part $\vec{q} = \vec{n} \sin \alpha$ of possible soliton fields of an electron. The left diagram corresponds to the choice $\vec{n} = \frac{\vec{r}}{r}$ for the internal colour degrees of freedom. The right diagram shows the configuration with $\vec{n} = (-\frac{x}{r},-\frac{y}{r},\frac{z}{r})$. Both soliton fields are equivalent since they differ by a global gauge transformation around the z-axis only.}
\label{electron}
\end{figure}

Schematic drawings of such stable soliton configurations are very instructive. In Fig.~\ref{electron}a we depict a field configuration with ${\cal Q} = 1$. The soliton field $Q$ is uniquely represented by its imaginary part $\vec{q} = \vec{n} \sin \alpha$. One can clearly see that the singularity of the $\vec{n}$-field at the center is suppressed by $\sin \alpha$ approaching zero. An example of this field after a global gauge transformation is shown in Fig~\ref{electron}b. It results from a $\pi$-rotation of the colour vectors around the z-axis.

\begin{figure}
\centerline{\scalebox{0.5}{\includegraphics{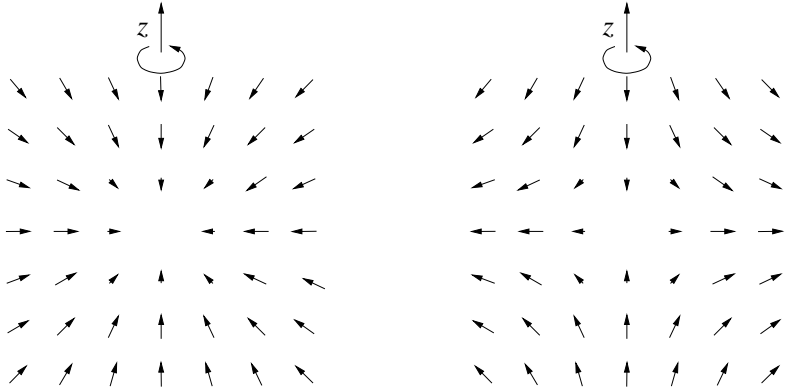}}}
\caption{Schematic diagrams of the imaginary part $\vec{q} = \vec{n} \sin \alpha$ of possible soliton fields of a positron. The left diagram corresponds to the choice $\vec{n} = - \frac{\vec{r}}{r}$ and the right diagram to $\vec{n} = (\frac{x}{r},\frac{y}{r},- \frac{z}{r})$ for the internal colour degrees of freedom. Both soliton fields are equivalent.}
\label{positron}
\end{figure}

For the choice $\vec{n} = - \vec{r}/r$ the soliton fields are characterised by negative topological and positive electric charge. Two such field configurations are shown in Fig.~\ref{positron}. The right diagram differs again from the left one by a rotation of the local coordinate systems by $\pi$ around the z-axis.

 Since negative winding numbers in Eq.~(\ref{Qreps}) are equivalent to negative $\vec{n}$ it is useful to consider the combined quantity $\vec{q} = \vec{n} \sin \alpha$ which is the imaginary part of the soliton field $Q$. In the left diagram of Fig.~\ref{electron} we show a schematic picture of the hedgehog solution (\ref{hedgehog}) with ${\cal Q} = 1$. By a global rotation of the locale coordinate systems we get other solutions with the same energy and topological charge since $\vec{\Gamma}$ transforms like a vector (\ref{PotgaugeTrafo}). The right diagram of Fig.~\ref{electron} shows the hedgehog solution after a $\pi$-rotation around the z-axis.

As we discussed already in Sect.~\ref{SecElecMonop} there seems to be a strong similiarity between the properties of solitons and charged leptons. One could therefore try to identify higher topological charges with $\mu$ and $\tau$ leptons. But it seems difficult to describe the high mass rations between these leptons.

Due to the potential energy term (\ref{potential}) we can't construct soliton states of finite energy with an even topological charge (\ref{uncharged}) and a profile function $\alpha(\vec{r})$ depending only on the radius $r$. We can ask whether soliton fields are possible with a profile function which is not spherically symmetric? Let's concentrate for symplicity at configurations which cover $S^3_{1/2}$ twice. For them $\alpha(\vec{r})$ would have to approach in all radial directions an element with zero potential energy, $Q(\infty,\vartheta, \varphi) = i \vec{\sigma} \vec{n}(\infty,\vartheta, \varphi)$. Somewhere in space one expects two center elements of $SU(2)$, which would also be the centers of two interacting solitons. Such configurations we are going to discuss in the next section. There we will also develop some ideas about the relation of such configurations to spin quantum numbers.

\subsection{Many-particle states}\label{ManyParticleStates}

There are some interesting puzzles in the conventional treatment of many-fermion systems. In quantum mechanics we expand states of many fermions in a basis of product wave functions, of Slater determinants. The antisymmetry of many particle wave function guarantees that two equal particles don't occupy the same place. When the interacting particles leave the interaction region they become independent and finally only single particle states are measured in a detector. Due to antisymmetry of the many-particle wave function quantum mechanics predicts in general an entangelment of states and can't predict which components are really measured. The reduction of the wave function in the measurement process is still an open question. A further puzzle concerns the non-locality of the anti-symmetric wave function and is presently again in hot discussion under the name of the Einstein-Podolsky-Rosen paradox.

In this article we treat fermions as topological excitations of a bosonic field. Because of their topological structure the Pauli principle in its first version, that fermions can't occupy the same quantum state, is guaranteed automatically. The simplest and very instructive example is a pair of solitons in the sine--Gordon-model. If two solitons approach each other, they are repelled. A two-soliton function is a function of one coordinate only describing a doubly charged state. It is not constructed as a product wave function of two essentially point-like objects. In analogy to this behaviour, there is no need for an antisymmetry of a fermionic many-particle state in our topological model, there is no reduction of the wave function after a spatial separation of scattering particles and no EPR-paradox for fermions. These effects seem to be connected with the description of a many-fermion state by a superposition of antisymmetric product wave functions of point-like structureless objects.

We will now discuss qualitatively some two-particle configurations in our soliton model. In this way we will also get some insight how the spin properties of fundamental particles may be implemented. The simplest two-particle states could be particle-antiparticle, positronium-like, states. In the sine--Gordon-model such states are not stationary. Solitons and antisolitons attract each other and annihilate. In a mathematical construction we can stabilize soliton-antisoliton solutions by appropriate constraints in a variational procedure. We constrain our solutions by fixing the positions of the centers of positive and negative charges to
\begin{equation}
\vec{R}_+ = (0,0,-\frac{R}{2}) , \quad \vec{R}_- = (0,0,+\frac{R}{2}) .
\end{equation}
and assume for simplicity the distance $R$ between the two centers to be much larger than $r_0$.

\begin{figure}[h]
\centerline{\scalebox{0.4}{\includegraphics{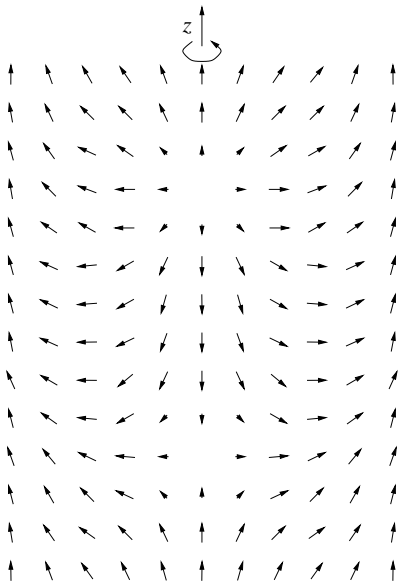}}}
\caption{Schematic diagram of the field $\vec{n} \sin \alpha$ for a positronium-like state. It is rotational symmetric arout the z-axis. We want to underline that the arrows do not indicate the electric flux lines but rather the values of the soliton field. It is interesting to see that approaching centers do not lead to an increase of curvature of the soliton field, the two opposite charges can easily annihilate if they belong to the same hemi--sphere of $S_3$ as discussed in the text.}
\label{FieldPositronium}
\end{figure}
\begin{figure}[h]
\centerline{\scalebox{0.4}{\includegraphics{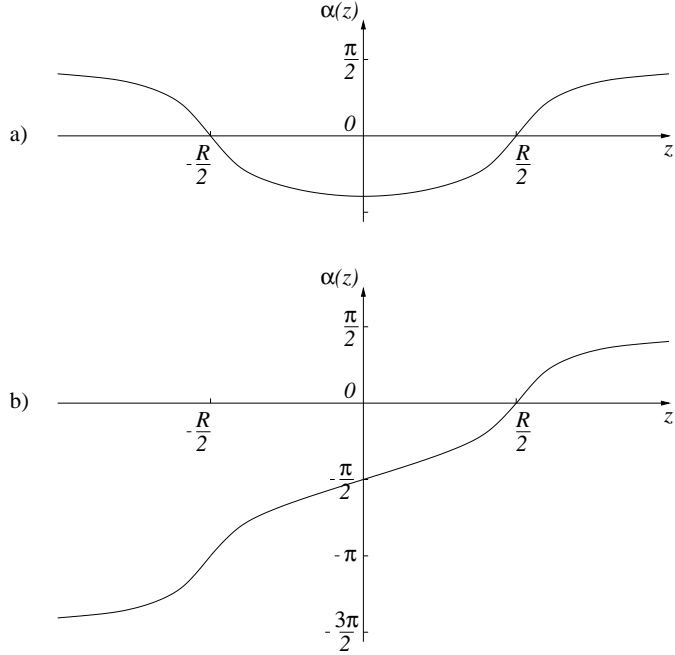}}}
\caption{Schematic profile of the behaviour of the internal variable $\alpha$ along the z-axis, $Q \; = \; \cos \alpha + i \;  \sigma_3 \sin \alpha$ is assumed. The upper diagram corresponds to a spin-singlet state and the lower to a spin-triplet state of a positronium like state.}
\label{SchemePositronium}
\end{figure}

Positronium is uncharged and the $U(1)$ gauge field gets trivial at infinity, i.e. it approaches a constant. Therefore, without loss of generality we can fix the soliton field at infinity to $Q(\infty)=i\sigma_3$. An appropriate combination of the fields of solitons and antisolitons of Fig.~\ref{electron} leads to a common soliton field of the type depicted in Fig.~\ref{FieldPositronium}. It is rotational symmetric around the z-axis. The quaternion field in the center of the positive and negative charge is a center element of $SU(2)$. It can still be chosen arbitrarily. It can be easily understood from Fig.~\ref{SchemePositronium} that configurations with equal center elements $Q(\frac{R}{2})=Q(-\frac{R}{2})$ have lower energy than those with different center elements $Q(\frac{R}{2})=-Q(-\frac{R}{2})$. Fig.~\ref{SchemePositronium} shows the schematic profile of the behaviour of the internal variable $\alpha$ along the z-axis. The upper diagram with two equal center elements has no zero at $r=0$ and corresponds to a quaternion field which covers one hemisphere of $S^3$ twice, at least for large distance of the two charges. 
For an approaching pair the  coverings of the hemisphere are not anymore complete and the curvature term in the energy should reduce. Such a configuration seems to correspond to the ground state of positronium which is a spin-singlet state. As first numerical calculations show the two charges feel an attractive Coulomb potential at large distances and start to annihilate at short distances. This can also be seen graphically from Fig.~\ref{FieldPositronium}. Approaching charges do not lead to an increase of the curvature of the soliton field. The lower diagram of Fig.~\ref{SchemePositronium} depicts a configuration which covers the whole sphere $S^3$ and can't contract to a trivial field. It could be assigned to triplet positronium which can only decay via emission of three photons.

If we assign the two positronium-like states to singlet and triplet states of positronium it may seem naturale to relate the center elements of solitons $Q=\pm 1$ to spin-up and spin-down states of electrons. But the relation between magnetic spin quantum numbers and center elements is not so simple. Quantum mechanics allows a slow rotation of electrons from spin-up to spin-down states. In this soliton model one can`t transform a soliton with a center element $Q=1$ by a rotation to $Q=-1$. This gives the hint that soliton fields which differ by a multiplication with a center element should be treated as identical. In other words the soliton configurations should be classified by the group $SO(3) = SU(2)/Z_2$. 
In this case one could not directly assing a magnetic spin quantum number to soliton fields modelling a single electron. Only to many--particle states one could assign a spin via the different possibilities to get connected field configurations in $SO(3)$. The two types of configurations discussed above would then correspond to two configurations on $S^3_{1/2}$ with differently connected double coverings of $S^3_{1/2}$. The singlet configuration with a trivial connectedness and the triplet configuration which is connected only via the identification of pairs of $SU(2)$ elements with the same element of $SO(3)$. In this sense there is not much difference between $SU(2)$ and $SO(3)$ configurations and we can continue to think in terms of $SU(2)$ field configurations which seem much simpler to imagine and to depict. We have to keep in mind only that field configuations which differ by a global multiplication with a center element are identical.

With the $SO(3)$ interpretation there remains the essential difference to quantum mechanics that to single electrons we can't attribute a spin quantum number. In this respect it may be interesting that there is no successful Stern-Gerlach experiment for single electrons. According to the knowledge of the author such types of experiments work only for uncharged systems, like the famous experiment which was done with neutral Ag-atoms. In our model a rotation of a single electron would correspond to a global gauge transformation (\ref{rotBasis}) of the soliton field. Such a transformation by $\pi$ around the z-axis was applied in Fig.~\ref{electron} between left and right diagram. It seems obvious that such a rigid rotation in the whole space is not possible in finite time due to the finite speed of light. Such a rotation seems easily possible for an uncharged Ag-atom whose soliton field should be non-trivial only inside the atomic radius.

\begin{figure}
\centerline{\scalebox{0.4}{\includegraphics{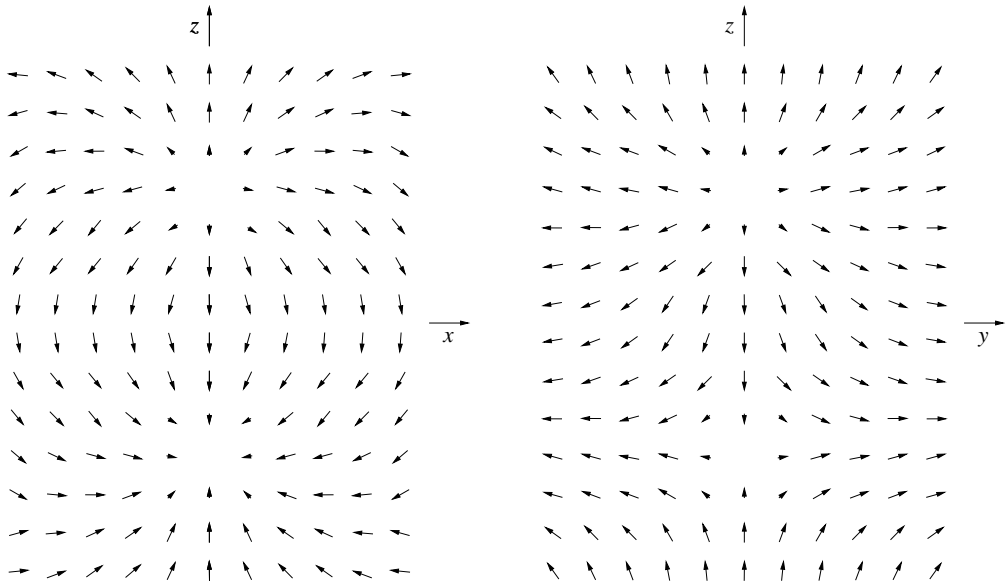}}}
\caption{Schematic diagram of two cuts through the field $\vec{q} = \vec{n} \sin \alpha$ for a doubly charged soliton configuration.}
\label{FieldTwoElectrons}
\end{figure}

Other interesting field-configurations are those corresponding to electron pairs. One can get a first trial configuration of that type by glueing together two identical soliton configurations, like that in the left diagram of Fig.~\ref{electron}. But in this way one gets a field configuration with very high curvature in the boundary plane, i.e. a highly excited state. In a minimisation process the two solitons start to rotate against each other and end up in a configuration which is schematically shown in Fig.~\ref{FieldTwoElectrons}. We want to emphasize that these diagrams depict the imaginary part $\vec{q} = \vec{n} \sin \alpha$ of the soliton field and not the electric flux lines. The electric flux lines are given by the curvature of the soliton field. Outside of the soliton cores at $\vec{r} = (0,0,\pm R/2)$ they are expected to be identical to the well-known flux lines between equal charges. In the soliton field in Fig.~\ref{FieldTwoElectrons} the part corresponding to the upper electron is identical to the soliton field in the left diagram of Fig.~\ref{electron}. The lower part results from a rotation by $\pi$ around the y-axis around the center of the corresponding soliton. 
One can clearly see that at large distances from the charge centers the soliton field approaches $\text{sin} \alpha = 1$ and covers the internal sphere $S^2$ twice as it is necessary for a system with electric charge $-2 e_0$. The two cuts through the soliton fields show that rotational symmetry is lost on the level of the soliton field. It should not be lost on the level of electric flux lines for distances large compared to $r_0$ after a gauge transformation of the $SU(2)$-field to a $U(1)$-field. As first numerical calculations show such field configurations have the behaviour of a repulsive Coulomb potential. This behaviour is expected also from the shape of the soliton field. If the two equal charges approach each other the curvature in the x-z-plane increases leading to a repulsion of the two charges. Again we expect an influence from the choice of the $SU(2)$ center elements at the two charge centers. For $Q(\frac{R}{2})=Q(-\frac{R}{2})=\pm 1$ the double covering of the hemisphere $S^3_{1/2}$ are not complete. This state is therefore slightly favored by the curvature energy and should describe the $^1 S_0$ ground state of a double electron system. The soliton field is in this case symmetric under a $\pi$-rotation around the y-axis which exchanges the two electrons. For the other case, $Q(\frac{R}{2})=-Q(-\frac{R}{2})$ two different hemispheres in internal space are completely covered corresponding most likely to an excited spin-triplet state. According to the above described $SO(3)$ interpretation one can extract the non-trivial center element of $SU(2)$ from the field of the corresponding electron, e.g. for negative $z$-values. This reverts the arrows in the negative $z$-region of the diagrams in Fig.~\ref{FieldTwoElectrons}. The soliton field is then asymmetric against an exchange of the two particles via a $\pi$-rotation around the y-axis.

This first qualitative analysis seems to indicate that one can possibly relate the choice of center elements in the centers of the solitons with the spin quantum numbers in quantum mechanics. A more detailed analysis including orbital angular momenta will of course be necessary for a deeper understanding of the relation of the soliton field to the quantum mechanical wave-function and its symmetry properties. In this respect it will also be of big interest to study the behaviour of single solitons and many-soliton configurations in external magnetic fields and to compare it with the well-known experimental results.

\section{Conclusions}

Our picture of the submicroscopic world has moved far away from intuition. There may be some hope that we can find a description which leads us to a deeper understanding of fundamental particles and their properties.

In this model we generalise the sine--Gordon model to four dimensions. Our aim is to treat fermions as topological excitations of a bosonic field. We introduce a field of unimodular quaternions, an $SU(2)$-field for the internal degrees of freedom. The Lagrangian contains essentially two terms, a curvature term and a potential energy term. The curvature term agrees with the Skyrme term of the Skyrme model, it corresponds further to the curvature term in gauge theories. With respect to gauge theories there is the essential difference that in gauge theories the basic field is the gauge field which defines the parallel transport of vector or spinor fields from one local coordinate system to another. In our model the basic field is the soliton field. The gauge field is a derived field, the field of connection coefficients between different local coordinate systems and can be expressed fully in terms of the soliton field. The second term in the Lagrangian is a potential term. It corresponds to the potential term in the sine--Gordon model and leads to the stabilisation of non--trivial soliton solutions. It breaks the rotational symmetry in the internal space $S^3$. Since the bottom of the potential is a two-dimensional manifold it allows for two types of massless excitations, the two polarisations of freely propagating electro-magnetic waves.

This model has several interesting features which solve long standing puzzles in particle physics.
\begin{itemize}
\item It explains by topology why only electric particles appear in nature and not their magnetic counterparts.
\item It treats fermions as extended objects. Therefore, many deep difficulties of quantum field theories seem not to appear in this theory. There seems to be no need for renormalisation leading to a momentum dependence of coupling constants. Such a dependence on the transfered momentum appears quite naturally for topological fermions as an analogon to Hooke's law for elastic bodies at a microscopic level.

\item The origin of mass is explained by the field energy of these extended particles.

\item Relativistic properties of fundamental particles, like velocity dependence of mass and Lorentz contraction, can be understood at a microscopic level analogous to the sine--Gordon model.

\item There is no need to introduce special anticommuting fields, Grassmann fields, to describe the dynamics of fermions. All field components are real valued in this model. The essential point in the description with quaternionic fields seems to be that unimodular quaternions allow for a smooth transfer of energy from one degree of freedom to another. Such a possibility for an energy transfer we find in many basic phenomena in physics, especially in all oscillatory motions. In oscillating pendula the energy oscillates between kinetic and potential energy, in two coupled pendula the energy is transfered periodically between the two pendula, in electromagnetic waves the field energy oscillates between electric and magnetic field components, in the quantum mechanical description of a free particle the energy oscillates between the real and the imaginary part of the wave function, etc. In our model two of the field components seem to be undamped, those are essentially the two polarisations of the electro-magnetic, of the $U(1)$-field. All other components are massive and therefore damped. They cannot propagate freely in space. This last property seems to be important for an extension of the model to other fundamental forces, like the colour and the weak force.

\item Our model of topological fermions suggests a ``classical'' resolution to the EPR paradox for fermions. The wave functions of many fermions are not antisymmetrised product wave functions of point-like particles but rather intrinsically of many body type. For large distances topological fermions approach  asymptotically unique single-particle states. Our model possibly sheds also some light on the problems which appear in quantum mechanics in connection with the reduction of the wave function in the measurement process.

\item There is a well-known doubling problem in the formulation of fermionic fields on a discretised space-time lattice. The soliton model of fermions resolves this problem by treating them as extended objects.
\end{itemize}

There remain of course many open questions in connection with this topological model of fermions. Some of them were already mentioned.

\begin{itemize}
\item It is not yet clear whether the running of the coupling, the dependence of the strength of the interaction on the transfered momentum, is described quantitatively correct.

\item A quantisation of the model is possible via the path integral formulation. Quantum corrections can be evaluated around the soliton solutions. A related problem concerns the analysis of the renormalisability of this quantised theory.

\item There are many questions in connection with the spin properties of solitons and their interaction with an external electro-magnetic field. Especially it should be clarified if the interaction of singlet and triplet positronium states with an external magnetic field is adequately described.

\item The structure of excited positronium states with non-zero angular momentum must be studied and the decay properties of positronium states have to be compared with the experimental predictions. In our model singlet positronium can annihilate without topological restrictions, since its winding number is zero. In the quantum field theoretical description it decays via two--photon emission, whereas triplet positronium decays via three-photon emission. In our model triplet positronium has winding number two. It's decay is inhibited and the appropriate mechanism for its decay is still unclear. 

\item We studied a special form of the potential energy density ${\cal H}_p$. There are other forms possible which lead to finite energy predictions. Since the prediction of the spectrum of charged leptons seems not to be in very good agreement with the experiment, one should possibly think about other dependences.

\item In potentials with infinitely high walls one expects excited states characterised by the number of osziallor quanta. Such states one could also expect for the solitons of this model. Are these the same states which we have characterised by the winding number $n_w$? It is an interesting question if such states could be excited by the transversal electromagnetic field.

\item A more general question concerns the extension of the model to higher groups treating other types of interactions. It seems quite clear that an extension to $SU(3)$ should be used to describe the colour force inside strongly interacting particles. According to the standard model, W-bosons are the gauge bosons of the weak interaction. One could speculate that W-bosons, as they have a definite mass, are not quanta of fundamental fields but rather composite objects, i.e. topological excitations of an $SU(4)$- or $SU(5)$-field.

\item Baryons in an $SU(3)$-model should most likely not be described by three separated fundamental objects, the quarks, rather by topological stable $SU(3)$ field configurations describing the whole baryon. This may lead to an explanation of confinement by intrinsic topological properties of the gluon field. There would be no need for a non-trivial vacuum, a picture which is used in Quantum Chromo Dynamics.

\item A deep puzzle concerns the structure of neutrinos. Originally the author believed that neutrinos are connected with even winding numbers of the quaternion field. But this idea turned out to be erroneous. It is not yet clear how neutrinos should be included in some version of this topological model of fermions.
\end{itemize}

If some version of this model should turn out to be a good candidate for the description of fundamental particles it should be compatible with Quantum Mechanics. Via the path integral formulation of Quantum Mechanics this should be easily possible.

The differential geometrical formulation of this model seems to give a natural possibility for the inclusion of general relativity. But it is not clear how a gravitational $1/r^2$-force could appear between neutral particles. Better understandable seems a variation of the speed of light in matter and a deflection of plane electro-magnetic waves by solitons.

\section{Acknowledgment}
I would like to thank Peter M\"obius for his enthusiastic presentation of the properties of sine--Gordon solitons with his mechanical transmission line of elastically coupled pendula at a conference in Smolenice in the year 1988 which initiated my interests in a four--dimensional generalisation. Further, I thank my wife and my children for their patience when I followed during many vacations and evenings wrong tracks of the above model. I am also thankful to Friedrich Manhart for his support in understanding basic concepts of differential geometry. This work was supported in part by Fonds zur F\"orderung der Wissenschaftlichen Forschung P11387-PHY.

\end{document}